\documentclass[10pt]{iopart}

\usepackage{iopams}  
\usepackage{cite}  
\usepackage{graphicx}
\newcommand{\angstrom}{\textup{\AA}}
\usepackage{amssymb}
\usepackage{units}

\newcommand{\db}[2][]{\text{d}^{#1}#2}

\newcommand{\text}[1]{\textnormal{#1}}

\newcommand{\abs}[1]{|#1|}
\newcommand{\al}{\ensuremath{\alpha}}
\newcommand{\be}{\ensuremath{\beta}}

\begin{document}

\title[Microscopic modeling of gas-surface scattering. II. Ar on Pt(111) surface]{Microscopic modeling of gas-surface scattering. II. Application to argon atom adsorption on a platinum (111) surface}


\author{A Filinov$^{1,2,3}$, M Bonitz$^1$ and  D Loffhagen$^2$}
\address{$^1$ Institut f\"ur Theoretische Physik und Astrophysik,
Christian-Albrechts-Universit\"at, Leibnizstr. 15, D-24098 Kiel, Germany}
\address{$^2$ INP Greifswald e.V., Felix-Hausdorff-Str.~2, D-17489 Greifswald, Germany}
\address{$^3$ Joint Institute for High Temperatures RAS, Izhorskaya Str.~13, 125412 Moscow, Russia}
\ead{filinov@theo-physik.uni-kiel.de}

\begin{abstract}

A new combination of first principle molecular dynamics (MD) simulations with a rate equation model presented in the preceding paper (paper I) is applied to analyze in detail the scattering of argon atoms from a platinum (111) surface. The 
combined 
model is based on a classification of
all atom trajectories according to their energies into trapped, quasi-trapped and scattering states. The number of particles in each of the three classes obeys coupled rate equations. The coefficients in the rate equations are the transition probabilities between these states which are obtained from MD simulations. While these rates are 
generally time-dependent, after a characteristic time scale $t_E$ of several tens of picoseconds they become stationary allowing for a rather simple analysis. Here, we investigate this time scale by analyzing in detail the temporal evolution of the energy distribution functions of the adsorbate atoms. We separately study the energy loss distribution function of the atoms and the distribution function of in-plane and perpendicular energy components. Further, we compute the sticking probability of argon atoms as a function of incident energy, angle and lattice temperature.
Our model is important for plasma-surface modeling as it allows to extend accurate simulations to longer time scales. 

\end{abstract}



\submitto{\PSST}                            

\maketitle

\ioptwocol

\section{Introduction}\label{s:intro2}

The recent progress in the production of micro- and nanoelectromechanical systems, aerospace engineering, vacuum technologies, and process engineering, such as gas separation membranes and heterogeneous catalysis, have revived the interest in fundamental research of rarefied gas flows and heat transport. 
Here, the influence of the gas-surface interactions on the momentum and energy transfer leading to the wall slip effects is of primary interest. The resulting macroscopic properties, such as the momentum/energy accommodation coefficients,  can then be efficiently used as the input parameters for scattering kernels in rarefied gas flow simulations and well-defined boundary conditions. 

Plasma-surface interaction is another currently actively studied field, where surface processes involving neutral atoms 
and molecules are of high importance. 
This includes the understanding of the film formation 
on the surface as well as feedback effects of the surface processes 
to the plasma composition and discharge characteristics. 
The questions of interest include the adsorption of gas atoms and molecules, their sticking probability, 
the generation of secondary particles, 
as well as the dependence of the adsorption on the surface temperature and the energy 
and impact angle of the 
gas particle. These properties are often known only approximately since both measurements and simulations in a plasma environment are quite complex. Although hydrodynamic modeling and kinetic  simulation 
of plasmas have progressed remarkably during the recent decade, the interaction of the plasma with a solid surface is typically treated on a simplified level via averaged sticking and desorption rates. Here, improvements, e.g. by obtaining energy resolved desorption rates,  
would allow for much more accurate modeling of low-temperature plasmas. In this paper we present new results for these quantities for the test case of argon atoms impacting on a platinum surface.  

In the last decades the scattering problem of rare gas atoms from metal surfaces continued to be in the focus of both theoretical~\cite{tr1,pol09} and experimental analysis~\cite{exp1}. A detailed overview on the recent progress in this field can be found in Refs.~\cite{rw1,rw2,rw3}. The analyses include both classical and quantum scattering regimes. In particular, it was found that some quantum features in the diffraction patterns can be resolved 
even for atoms as heavy as Ar~\cite{qs}. 
Such a situation, however, is quite unusual and most of the scattering experiments with heavy atoms are of  classical nature~\cite{cl1,cl2,cl3}. But even in this case, a classical mechanics picture of the atom-surface scattering  is far from being complete and further developments are of  interest. 

There are several main problems for a theory. The first one is an accurate inclusion of the gas-surface interaction, typically reconstructed from ab initio quantum mechanical approaches such as 
density functional theory (DFT). 
A second issue concerns  
a correct consideration of all energy dissipation channels, like the interaction 
of the gas 
with the surface and bulk phonons, and  electron-hole excitations originating from the perturbation of the electron density of the surface atoms. These major processes have to be taken into account by a theory to  describe  correctly the energy transfer to the surface, the angle- and energy-dependent sticking probability, the equilibration kinetics of the adsorbate localized at the surface, and the subsequent thermal 
desorption,  
which extends over a time scale ranging from  few picoseconds to milliseconds. An important test for a theoretical description is certainly provided by  experimental measurements performed on different physical systems and scattering conditions.  

Concerning experimental studies, the 
 dependence of the angular distribution of scattered atoms on  
 the incident energy of the gas atoms and the surface temperature 
 as well as the average energy loss to the surface are typically in the focus. The latter, being the most challenging to theory, is typically defined via 
  the so-called accommodation coefficient, i.e., the ratio of the energy loss to the incident energy of the gas particles. In particular, theories, which assume parallel momentum conservation and hard sphere scattering, are not able to capture the main experimental observation that the slope of the relative final energy changes from negative to positive when increasing the incident energy of the gas particle. This behavior has been observed 
  e.g.\, for the scattering of Ar atoms and diatomic molecules (N$_2$ and O$_2$) from a Ag(111) surface~\cite{rau} and for the scattering 
  of Xe atoms from a Pt(111) surface~\cite{rt1}.  

A real theoretical breakthrough in the understanding of experimental data has been gained by using molecular dynamics (MD) simulations. The experimental results on scattering of Xe on Pt(111) have been reproduced by Barker \etal~\cite{md1}. The energy loss distribution function and its dependence 
on the incident energy and angle has been accurately analyzed by Lahaye \etal~\cite{Lah} for Ar on Ag(111). 

While being in good agreement with experimental data, 
MD simulations alone cannot 
always provide a deep understanding of the underlying physics. Therefore,  several theoretical models of scattering have been developed in parallel. In general, they treat the interaction 
of the gas atoms 
with the surface 
by coupling  their vertical motion to the phonon bath and use the methods of the classical statistical physics. In this framework, Brako~\cite{brako} obtained an analytical expression for the energy and momentum transfer as well as for the angular distribution. However, the corrugation of the surface was neglected 
for reasons of simplification. As a result, Brako's theory misses the coupling between the phonons and the parallel momentum of the atom. 

The first model which took this effect into account -- the so-called  ``washboard model'' -- was developed by Tully~\cite{tully90} and was    extended in a later paper by Yan \etal~\cite{Yan}. 
It assumes an impulsive collision of the gas atoms  with the surface, but shows difficulties in reproducing the double peak structure of the angular distribution function. 

The first progress in this direction was made by Pollak \etal~\cite{pol09}. They generalized  Brako's formula and the ``washboard model'' and presented explicit results for the joint angular and 
final momentum distributions, the joint distributions of final scattering energy and angle as well as the interrelationship between the average energy losses and the angular scattering distribution. The application of this theory was successful in reproducing the main experimental observations. A shift of the maximum of the angular distribution to subspecular angles at low incident energies 
was predicted 
which is due to large energy losses in the horizontal direction. In contrast, a superspecular maximum in the angular distribution was reproduced 
at high incident energies, when the energy loss  in the vertical direction dominates. 
The theory predicts as well 
that the full width at half maximum of the angular distribution varies as the square root of the temperature. 
 It was successfully applied to explain the scattering experiments of Ar on the Ag(111) surface, however, 
 only in the regime when the incident energy is sufficiently large and one can neglect the sticking probability.              

The nature of sticking of heavy atoms to clean surfaces has been intensively investigated using semiclassical perturbation theory~\cite{pol09,hub,Mans,pol2011,pol2014,pol2015}. Hubbard and Miller~\cite{hub} calculated the sticking probability for the He–W(110) and Ne–W(110) systems in the regime when the energy transfer from the surface is small. Pollak\cite{pol2011} derived an expression for the sticking probability in the limit of weak surface corrugation, and by assuming weak coupling to the harmonic surface phonon modes. Later, an improved theory has been presented in~\cite{pol2014}, 
which included the second order corrections to the angular distribution of the scattered particles. In their recent work, Sahoo and Pollak~\cite{pol2015} employed a one-dimensional generalized Langevin equation and derived an analytic expression for the temperature-dependent energy loss. A combination with the multiple collision theory of Fan and Manson~\cite{Mans} allowed to determine the fraction of trapped particles after subsequent collisions (bounces) with the surface. The theory has been tested by comparison with  numerical simulations for the scattering of Ar on a LiF(100) surface.

In contrast to the classical theory, a quantum mechanical formalism capable of describing the scattering, trapping, sticking, and desorption for light particles, like He and H, has been also extensively studied, see Refs.\cite{q1,q2,q3} and references therein.

The effect of sticking and thermalization with the surface becomes increasingly important for surfaces covered with self-assembled monolayers consisting of long-chain functionalized molecules. This type of problem has been analyzed recently by Castej\'{o}n \etal~\cite{cast}. It was demonstrated how the efficiency of the energy exchange and the sticking probability vary with the length of the molecules in the monolayer. Longer molecules lead to an increased surface corrugation and provide an additional dissipation channel that promotes more efficient momentum and energy accommodation, and enhanced trapping.

Another aspect, which called for a systematic investigation, is 
the consideration 
of internal degrees of freedom of the scattering projectiles, such as the  molecules N$_2$ or CO$_2$. Following the statistical description, one samples the rotational energy of molecules according to the Boltzmann distribution~\cite{bbook},  
where the thermal occupation of the excited states is defined by the quantum number of the angular momenta operator and the rotational partition function~\cite{bbook1}. Then, the rotational energy is splitted between two rotational degrees of freedom to describe the rotation of the molecular around its symmetry axis~\cite{bbook1}. 
Moreover, the incident azimuth and altitude angles of the molecular bonds are randomized according to the specified incidence conditions. 
In this case, the scattering from the surface can bring a molecule to an excited rotational state and, thus, can constitute an additional dissipation channel~\cite{Li}.         

While direct MD simulations provide a detailed description of the adsorbate kinetics near the surface, such simulations are very time consuming 
in general. In order to overcome the restriction to relatively short time scales of the order of \unit[100]{ps}, we have developed a new approach 
for the modeling of the atom-surface interaction, which  is based on a 
combination of MD simulations with a 
rate equation model and which has been presented in detail in paper~I \cite{paper1}. There, we demonstrated how MD simulations can be used to reconstruct the quasi-stationary transition rates for the desorption processes on a much longer time scale. The basic assumption used was that the thermalization of the incident atoms with thermal and subthermal energy and the phonon-bath of the surface takes place on a time scale of the order of 50 to 
$\unit[100]{ps}$. 

In the present paper, we provide an additional confirmation of this assumption by performing a detailed analysis of the convergence of the energy distribution functions. The focus is on Ar atom scattering from a Pt(111) surface, which represents a quite well studied system. Mullins \etal~\cite{expPt} have conducted  detailed experimental studies on such system and reported on the changing role of the parallel momentum, $p^{\parallel}$, with surface temperature, $T_s$, for the trapping probability. The degree to which the parallel momentum is involved in the trapping process is frequently expressed by an energy scaling factor, $\cos^n \theta$, 
where $\theta$ is the angle of incidence of the gas atoms impinging onto the surface. 
The ``total'' energy scaling corresponds to $n = 0$, while the ``normal'' scaling corresponds to $n = 2$. The normal  scaling implies that the trapping probability is a function  of the normal momentum, $p^{\perp}$, only.
However,  the experiment showed a gradual change of the scaling factor with  increasing surface temperature from $n=1.5$ at  $T_s=\unit[80]{K}$ to  $n=0.5$ at $T_s=\unit[300]{K}$. It was suggested that high surface temperatures increase the surface roughness  
and  that 
parallel momentum dissipation also becomes increasingly more important to the trapping dynamics as $T_s$ increases. 

Such deviations from the normal energy scaling have been intensively investigated by means of  
numerical simulation~\cite{Tully,Tully2,smith1,gordon,smith2}. Both low and high temperature regimes have been analyzed. It was found that $\langle p^{\parallel}\rangle(t)$ shows a much slower convergence with time in comparison with $\langle p^{\perp}\rangle(t)$. This fact points to an incomplete equilibration of the parallel velocity prior to desorption. The desorbed atoms retain a memory of their incidence conditions, and this information enters into 
the desorption probability. 
Correlations with initial conditions retained during the dynamics of adatoms across the surface can be studied in depth in the simulations by monitoring 
the time dependence of the energy and other quantities. 
This approach is followed in the present studies as well. In particular, we look in detail how the energy exchange (or the energy-loss function) depends on the incidence conditions: the initial momentum, energy and lattice temperature. The main non-adiabatic process is the exchange with lattice phonons.

The present work extends the earlier analyses~\cite{Tully,Tully2,smith1,gordon,smith2,Leonard,Spijker} on the scattering of Ar on Pt(111) 
surfaces by including the effects of multiple reflections. 
This allows to resolve characteristic times 
for the convergence of the energy and energy-loss distribution function to a quasi-stationary state.

The paper is structured as follows. In section~\ref{energy} we analyze the temporal evolution of the  energy distribution 
of the adsorbate. 
The relaxation of the surface-normal and tangential energy components is analyzed in section~\ref{SecE}. 
Section~\ref{stick} presents results for the sticking coefficient of argon atoms on Pt(111) surfaces. 
It includes an analysis of its dependence on the incident energy and angle, and on the lattice temperature as well as a comparison with 
experimental data available from literature. 
Finally, a discussion and conclusions are given in section~\ref{con}.

\section{Evolution of the energy distribution of trapped and scattered adsorbate atoms}\label{energy}
  
In this section we analyze in detail how the system evolves towards equilibrium by analyzing the convergence to an equilibrium energy distribution function, and reveal possible dependencies on the incident angle ($\theta$), incident energy ($E_i$) and lattice temperature ($T_s$). 
We study mono-energetic gas atoms that 
are introduced at a height $z$ of 
\unit[20]{\angstrom} 
above the surface and  
are incident on a surface with $\theta[^\circ]=0 (30,60)$ and 
$E_i/(k_BT_r/e_0)=0.5 (0.62, 1.41)$ with 
 $k_BT_r/e_0=\unit[25.7]{meV}$ 
corresponding to the room temperature $T_r= \unit[300]{K}$. 
Here, the angle $\theta=0^\circ$ denotes normal incidence. 
The energies are chosen to maintain 
a similar value of the initial sticking probability in every case.

Argon atoms are trapped near the surface within the distance 
$z^{c}=r^c_{\mathrm{Ar-Pt}} \leq \unit[10]{\angstrom}$, 
where 
$r^c_{\text{Ar-Pt}}$ is the cut-off radius of the potential~\cite{paper1}. 
Their trajectories are followed up to $n_b=40$ bounces or until the atoms are scattered into the continuum, i.e., an atom can freely leave the surface region 
for heights $z(t)> z^{c}$. 
For each set 
of incidence conditions
$\{\theta,E_i,T_s\}$, 
the measured distribution functions, their mean values and variances are evaluated for a statistical ensemble containing $1000-5000$ trajectories.

When a gas atom interacts with the surface for the first time, 
the exchanged energy  depends 
crucially 
on the incident energy, but only to a lesser extent on the lattice temperature $T_s$. This was confirmed by previous studies 
of Smith \etal~\cite{smith2}, where it was shown that the mean energy loss is insensitive to $T_s$, while the width of the energy-loss distribution function increases with $T_s$. These observations are re-examined below with full temporal resolution to identify the presence of a quasi-equilibrium phase and to justify 
the introduction of the equilibration time $t^E$ and the equilibrium transition rates $T_{\al\be}^E$ 
according to paper~I~\cite{paper1}.

 \begin{figure}
  \begin{center} 
  \hspace{-0.0cm}\includegraphics[width=0.47\textwidth]{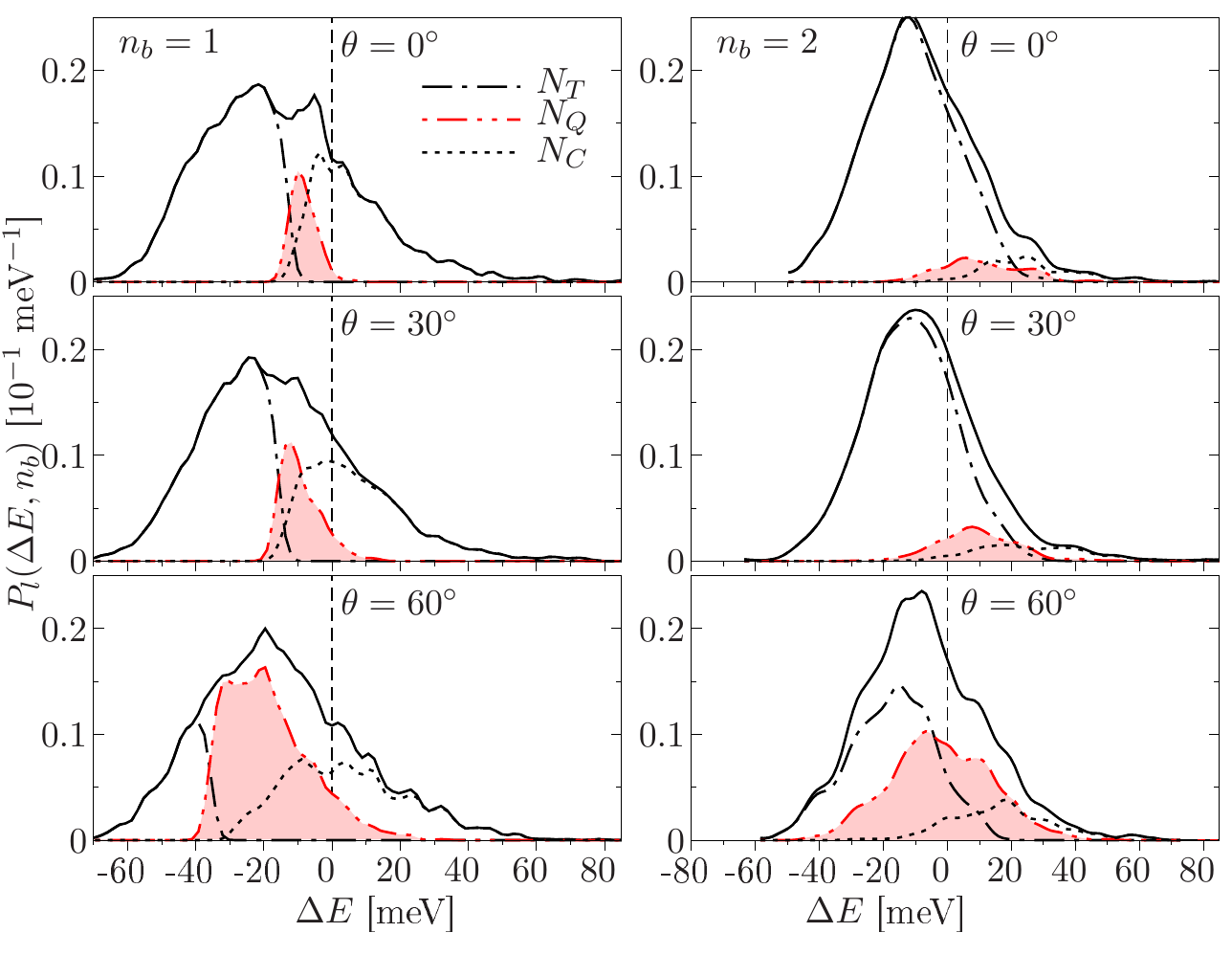} 
  \end{center}
  \vspace{-.80cm}
  \caption{Energy loss distribution $P_{l}(\Delta E; n_b)$  with $\Delta E=E_f-E_i$ 
  after $n_b=1$ (left column) and $n_b=2$ (right column) bounces at 
  the lattice temperature $T_s=\unit[190]{K}$. The contribution of the trapped, quasi-trapped (filled area) and continuum states is indicated for comparison.  
  Incident angles are $\theta=0^\circ(30^\circ,60^\circ)$ and the corresponding incident kinetic energies are chosen as 
  $E_{i}/(k_B T_r/e_0)=0.5 (0.62, 1.41)$  with $T_r = \unit[300]{K}$ corresponding to  $ k_B T_r/e_0 =\unit[25.7]{meV}$.}
  \label{fig:dEfiT06n1}
 \end{figure}

\subsection{Dynamics of the energy loss distribution function}
We start our discussion with the energy loss distribution function (ELDF), 
which is analyzed after a 
different number of bounces $n_b$ of argon atoms 
with the Pt(111) surface. 
Figure~\ref{fig:dEfiT06n1} compares the results for 
the three incident angles $\theta = 0^\circ$, $30^\circ$, and $60^\circ$ at $T_s=\unit[190]{K}$ after one and two bounces. 
Here,  
$\Delta E=E_f-E_i$ is the energy change during a single bounce event with the final ($E_f$) and incident ($E_i$) energy, 
$\theta=0^\circ$ corresponds to normal incidence, 
and $N_\mathrm{T}$, $N_\mathrm{Q}$ and $N_\mathrm{C}$ denote the fraction  of atoms in  trapped (T),  quasi-trapped  (Q)  and
continuum (C) states, respectively.  
The energy losses due to electronic friction are of minor importance in the present analysis, since the vibrational frequencies of Ar on the surface are much lower 
than the Debye frequency of the 
platinum surface. Therefore, the coupling to the  high frequency electronic motion should be negligible. 
We have tested this by explicit inclusion of a spatially varying electronic friction term in the equations of motion and 
found that the corresponding energy losses are below few meV. 
Notice that this effect becomes important in the high-energy regime  with incident energies exceeding $\unit[0.5]{eV}$. 

Figure~\ref{fig:dEfiT06n1} illustrates that the ELDF always extends from approximately $-70$ to $\unit[60]{meV}$ at the conditions considered. 
It has a similar shape 
 in all three cases 
 for $n_b=1$ and 2, respectively. The corresponding 
 mean value of the distributions is close to 
$\unit[-20]{meV}$ for $n_b=1$ and $\unit[-10]{meV}$ for $n_b=2$,   
and it does not depend on the incident angle  $\theta$. 
 However, the relative contributions of the different states, 
 i.e., 
 the T, Q, and C 
states, 
 is very different. 
For angles around $\theta=60^\circ$, a large fraction of initial energy is accumulated in 
the parallel kinetic energy 
$E_{i}^{\parallel} = E_{i} \sin \theta$. 
Hence, 
in order to bring a particle to a trapped state, 
i.e., a final state with $E_f = E_{f}^{\perp}+E_{f}^{\parallel}+V<0$,  
the parallel momentum should be significantly perturbed by the lattice: 
$\Delta E\propto (E_{f}^{\parallel}-E_{i}^{\parallel})\propto -E_{i}^{\parallel}$. 
Here,  
$E_{f}^{\perp}$, $E_{f}^{\parallel}$, and 
$V$ denote the perpendicular and parallel final kinetic energy and the trapping potential of the surface, respectively, 

We also observe that the value $(E_{f}^{\perp}+V)$ 
is similar in the three cases after the first reflection 
due to a similar value 
$E_{i}^{\perp} = E_{i} \cos \theta$ 
of the incident gas atoms. 
Hence, the energy loss of the trapped states increases with 
growing 
parallel 
incident 
kinetic energy 
$E_{i}^{\parallel}$.
This explains the shift of the maximum in the ELDF of the trapped states to negative energies with increasing~$\theta$. 
As a result,  the relative contribution of the T states in the full distribution function is reduced and 
taken over by the Q states. 
This trend can be directly followed on the left side of figure~\ref{fig:dEfiT06n1}, where the Q states are shown by the filled pattern.

  \begin{figure}
  \begin{center} 
  \hspace{-0.0cm}\includegraphics[width=0.47\textwidth]{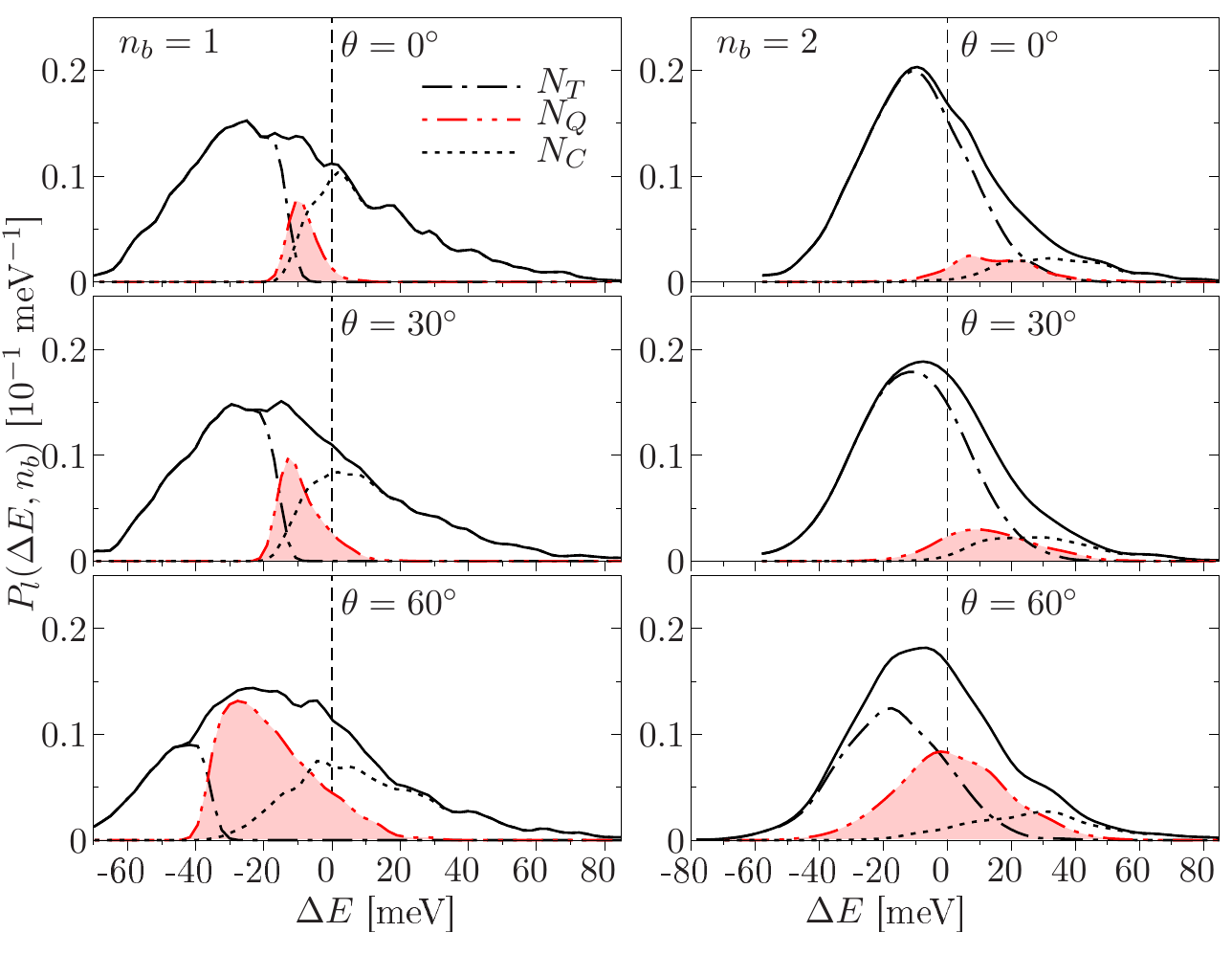} 
  \end{center}
  \vspace{-0.8cm}
  \caption{Same as in figure~\ref{fig:dEfiT06n1}, but for the lattice temperature $T_s=\unit[300]{K}$. 
  }
  \label{fig:dEfiT10n1}
 \end{figure}

The right column of figure~\ref{fig:dEfiT06n1} shows the ELDF after the second bounce. 
The contribution of the T states dominates for small incident angles ($\theta=0^\circ,30^\circ$). 
This results from the 
large 
transition rate $T_{TQ}$ 
leading to a fast conversion from Q to T states   
(cf.\ figure~9 in paper~I~\cite{paper1}).
For 
larger incident angles 
($\theta=60^\circ$), the population of the Q state becomes comparable to that of the T state for $n_b=2$ 
due to the large fraction of $N_\mathrm{Q}$ for $n_b=1$ being reduced by the second bounce.

From this first analysis, we can expect that the convergence to the 
quasi-equilibrium, 
i.e., when the T states dominate,  
takes longer for larger incident angles, 
because more bounces are required to reach such condition. 
In order to have  a similar initial sticking coefficient, 
the  energy $E_i$ in the incident beam has to be increased with 
increasing  angle of incidence $\theta$ due to the $E_{i}^{\parallel}$ component. 
This in turn increases the initial population of the quasi-trapped states 
and leads to a slower convergence to a quasi-stationary distribution.

The corresponding ELDF obtained for the larger lattice temperature $T_s=\unit[300]{K}$ 
is represented in figure~\ref{fig:dEfiT10n1}. 
The interpretation of these results is similar as for $T_s=\unit[190]{K}$ (figure~\ref{fig:dEfiT06n1}).  
 The main difference is the thermal broadening of the ELDF occurring at all angles. 

\begin{figure}
  \begin{center} 
  \hspace{-0.05cm}\includegraphics[width=0.47\textwidth]{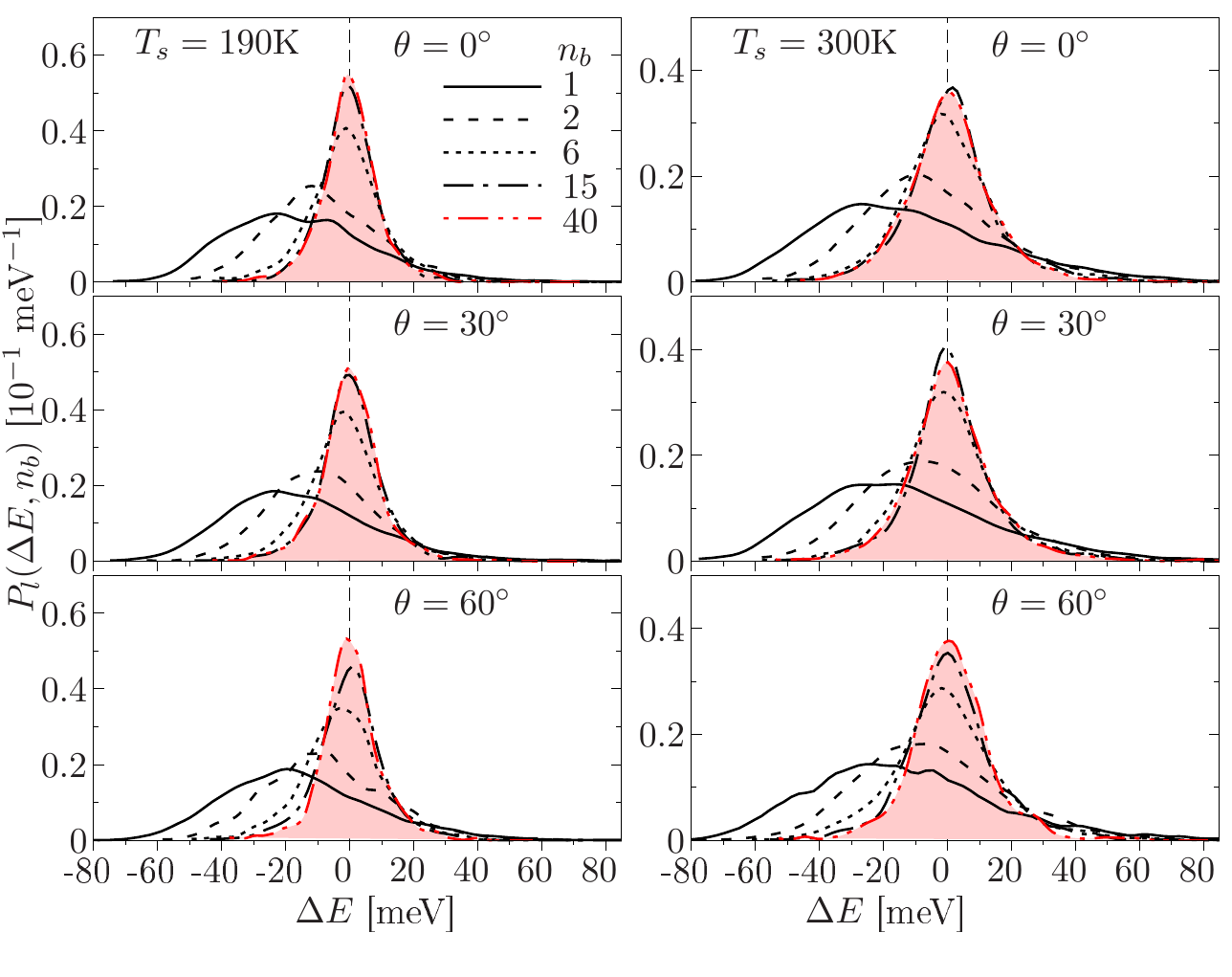} 
  \end{center}
  \vspace{-0.8cm}
  \caption{Energy loss distribution $P_l(\Delta E; n_b)$ for different number $n_b$ of  
  bounces with $1\leq n_b \leq 40$ at different 
  incident angles $\theta$ and  $T_s=\unit[190]{K}$ (left) and $T_s=\unit[300]{K}$ (right). 
  The converged distribution ($n_b=40$) is shown by the filled pattern.}
  \label{fig:dEfidiffn}
 \end{figure}

Figure~\ref{fig:dEfidiffn} shows the ELDF 
for different number of bounces $n_b$ 
at the incident angles $\theta = 0^\circ$, $30^\circ$, and $60^\circ$ and $T_s= 190$ and $\unit[300]{K}$.  
Because more bounces require longer time scales, 
this figure represents the  temporal evolution of the ELDF, 
where the correlation between average number of bounces and the absolute time scale 
is displayed in figure~7 of paper~I~\cite{paper1}.   
The results shown in figure~\ref{fig:dEfidiffn} clearly demonstrate that the adsorbate atoms actually equilibrate.  
The converged distribution is always reached after  $n_b=40$ bounces. 
It has a symmetric Gaussian-like form with respect to the absolute value of $\Delta E$. 
Here, the probabilities of excitation and de-excitation due to an absorption or emission of lattice phonons become equal. 

The rate of convergence of the ELDF to the stationary form 
depends on the incident angle. 
The distribution is close to the converged result 
 after about 15~bounces 
at both lattice temperatures 
for $\theta=0^\circ$ and $30^\circ$. 
In contrast, there are still some deviations from the stationary distribution 
for $\theta=60^\circ$, even for 
$n_b=15$. Here, full convergence is reached only for $n_b \geq 30$. 
The slower convergence at larger $\theta$ is directly related to the non-negligible contribution of the quasi-trapped states. 

The comparison of the results for the 
two lattice temperatures 
clearly shows that 
the effect of thermal broadening becomes larger with increasing  $T_s$. 
A corresponding quantitative analysis is presented in table~\ref{sigt}, 
where the variance $\sigma_l(\theta,T_s)$ of the ELDF
is given for the three incident angles and three lattice temperatures. 
The variance is obtained by fitting the ELDF 
for $n_b=40$ 
to the 
Gaussian form $e^{-(\Delta E)^2/(2 \sigma_l^2)}$. It 
exhibits a linear scaling with $T_s$ and shows  
no noticeable dependence on $\theta$. 
This justifies that the fitted distributions 
have a quasi-stationary form defined purely by the lattice properties, 
whereas any correlations with the incidence conditions ($\theta,E_i$) are practically lost.

\begin{table}[h]
\caption{Variance $\sigma_l(\theta,T_s)$ [meV] of the ELDF for different incident angles $\theta$ and lattice temperatures $T_s$. 
For $\theta= 0^\circ$, $30^\circ$, and $60^\circ$ the incident gas atoms have the initial energy $E_i/(k_BT_r/e_0)=0.5,  0.62$, and $1.41$, respectively.}
  \label{sigt}
 \begin{tabular}{c  c  c  c}
 \hline
 \hline
  $T_s$ (K) & $\theta=0^\circ$ & $\theta=30^\circ$ & $\theta=60^\circ$ \\
 \hline
 80 & 3.50(4)  & 3.48(4) & 3.60(6) \\
 190 & 7.20(10)  & 7.48(12) & 7.22(10)  \\
 300 & 10.13(24)  & 9.83(19) & 9.77(14)  \\
 \hline
 \hline
 \end{tabular}
 \end{table}
%

\subsection{Dynamics of the total energy distribution function 
of the adsorbate atoms}
 \begin{figure}
  \begin{center} 
  \hspace{-0.0cm}\includegraphics[width=0.46\textwidth]{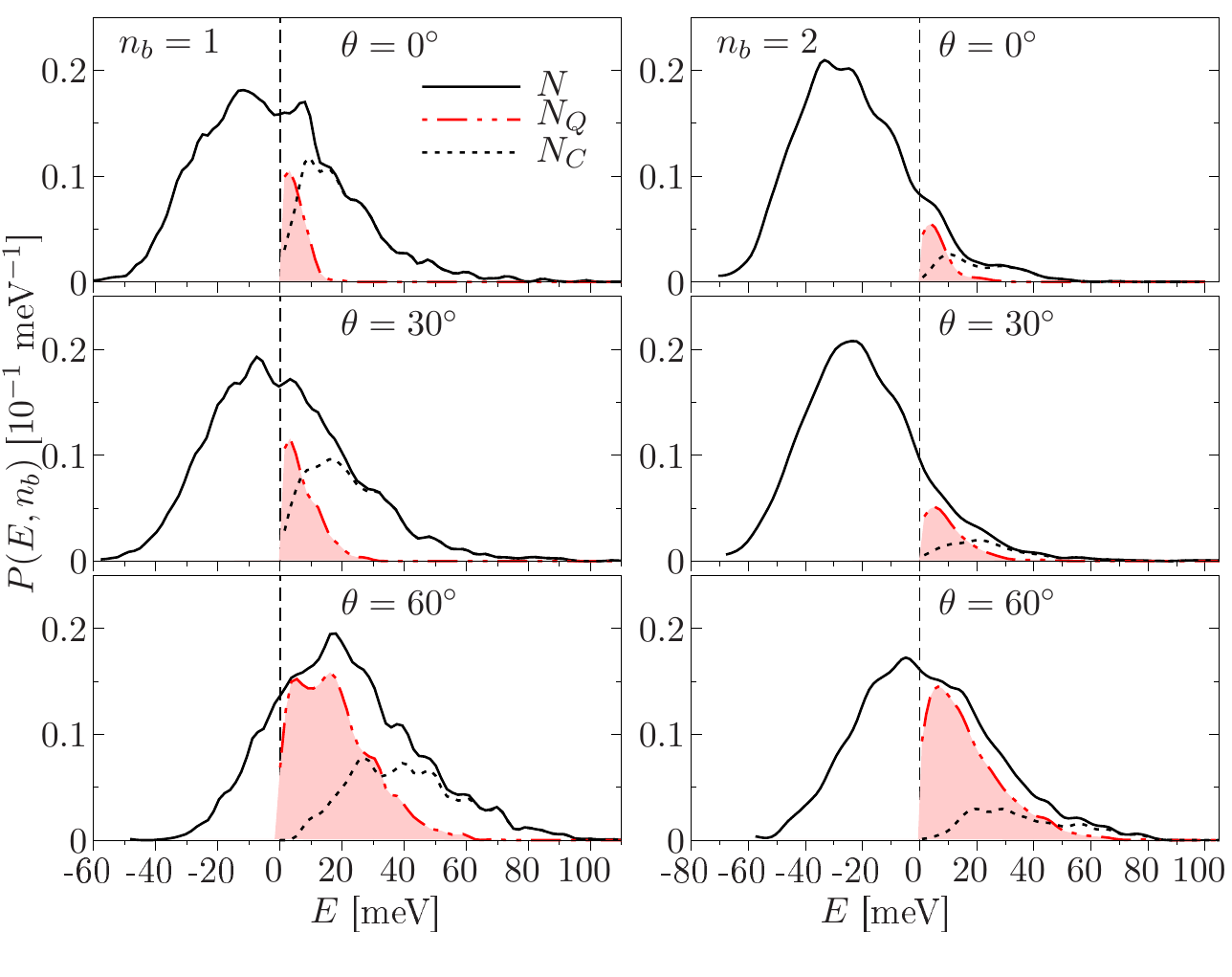} 
  \end{center}
  \vspace{-0.8cm}
  \caption{Total energy distribution $P(E,n_b)$ after $n_b=1$ (left column) and $n_b=2$ (right column) bounces 
  at 
  $T_s=\unit[190]{K}$ and incidence conditions as in figure~\ref{fig:dEfiT06n1}. 
  The contribution of the trapped states corresponds to $E< 0$. The quasi-trapped ($N_{Q}$) and continuum ($N_C$) states contribute for $E \geq 0$.  }
  \label{fig:Efnb12T190}
  \end{figure} 
  
 \begin{figure}
  \begin{center} 
  \hspace{-0.0cm}\includegraphics[width=0.46\textwidth]{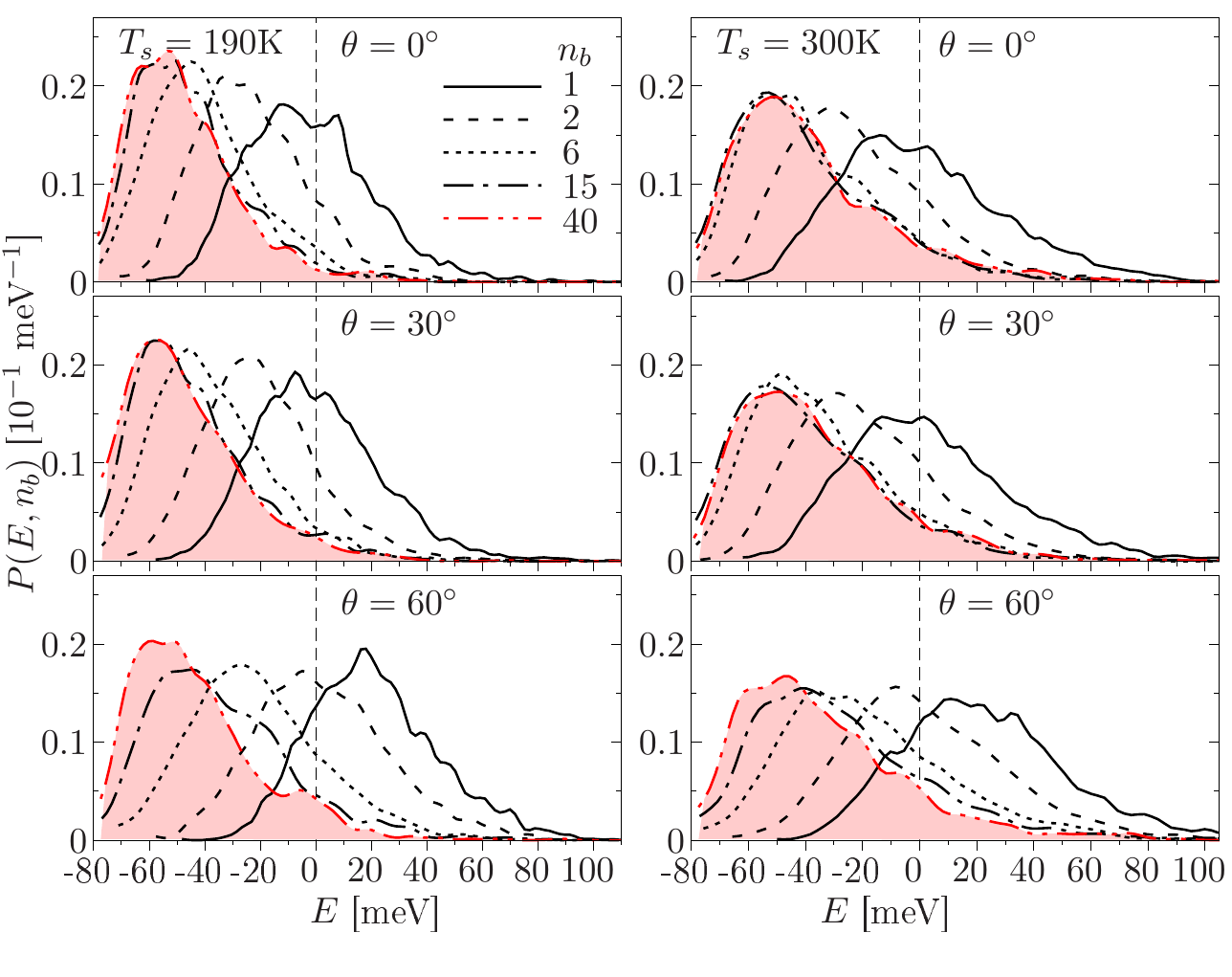} 
  \end{center}
  \vspace{-0.8cm}
  \caption{Total energy distribution $P(E, n_b)$  for different number $n_b$ of  
  bounces with $1\leq n_b \leq 40$ at 
  different 
  incident angles $\theta$ and  
  $T_s=\unit[190]{K}$ (left) and $T_s=\unit[300]{K}$ (right). 
  The converged distribution ($n_b=40$) is shown by the filled pattern.}
  \label{fig:Efnb12}
 \end{figure}  

The bounces lead to a change of the total energy distribution function (EDF) of the adsorbate atoms 
$P(E,n_b)$, which is represented in the figures~\ref{fig:Efnb12T190} and \ref{fig:Efnb12}.  
Figure~\ref{fig:Efnb12T190} presents the EDF of the 
adsorbate atoms after one and two bounces at 
the lattice temperature of \unit[190]{K}, which  
corresponds to the ELDF displayed in figure~\ref{fig:dEfiT06n1}. 
The EDF is composed of trapped, quasi-trapped and continuum states, 
where the contribution of the T states corresponds to energies $E< 0$ and 
the Q and C states contribute to the EDF for $E \ge 0$. 

Figure~\ref{fig:Efnb12} displays the EDF for different number of bounces $n_b \le 40$ at 
$T_s= 190$ and $\unit[300]{K}$, corresponding to the ELDF in figure~\ref{fig:dEfidiffn} and 
covering the entire time scale of the MD simulations. 
As before, we clearly observe that the convergence 
to  the
stationary distribution 
depends on the incident angle and that it should be explained by the initial population of the quasi-trapped states. 
We also find that the converged distribution 
corresponds mainly to the trapped states with the negative energies 
comparable to the depth of the physisorption well ($E_0\approx \unit[-78]{meV}$). 

Next, we analyze the evolution of the high-energy tail of the distribution functions. 
The tail with $E \ge 0$ is due to the quasi-trapped and continuum states. 
The fraction of these states typically decreases with 
increasing $n_b$ (or time). 
The  lower or higher energies are mainly due to the Q and C states, respectively. 
Note that the contribution of the Q states vanishes above some characteristic energy 
$E_c$, which is  
related to the lattice temperature according to 
$E_c=\nu \cdot k_B T_s$  with $\nu=2\ldots 3$. 
This effect can be clearly seen in figure~\ref{fig:Efnb12}. For instance, the high-energy tail of the EDF at $n_b=40$  
strongly decays above $\unit[40]{meV}$ 
for $T_s=\unit[190]{K}$ corresponding to $\unit[16.4]{meV}$, 
and  it extends  up to $60$ to $\unit[70]{meV}$ at the higher temperature 
$T_s=\unit[300]{K}$, which corresponds to $\unit[25.7]{meV}$. 

The reason, why the  contribution of the Q states can be observed in the EDF even in the regime of quasi-equilibrium ($n_b\gtrsim 30$), is directly related to the detailed balance condition derived in paper I~\cite{paper1}
\begin{eqnarray}
 N_T(t) T_{QT}(t)\approx N_Q(t)[T_{CQ}(t)+T_{TQ}(t)] \, ,
 \label{dbc2}
\end{eqnarray}
 which holds between the T  and Q states. 
 Here, $T_{QT}$, $T_{CQ}$, and $T_{TQ}$ denote the 
 transition rates from T to Q, from Q to C, and from Q to T states, respectively. 
 A finite but relatively small fraction of the Q states is sustained by the excitation channel  $\mathrm{T} \rightarrow \mathrm{Q}$. 
 By analyzing our MD data we found that this situation is common for high lattice temperatures 
 ($T_s=190$ and $\unit[300]{K}$), see $N_Q(t)$ in figures~5 and~7  of paper I~\cite{paper1}. 
 The same effect is also present for the lower $T_s=\unit[80]{K}$ when the thermal desorption is strongly suppressed. 
 It can be resolved from the curve $N_Q(t)$ in figure~3 of paper I when plotted in a logarithmic scale. 

This behavior can be understood from the following arguments. 
If the desorption of the trapped states to the continuum is most probable via a two-stage excitation, 
i.e., $\mathrm{T}\rightarrow \mathrm{Q}\rightarrow \mathrm{C}$, 
the quasi-trapped states should be present as a key ingredient of the desorption kinetics. In this case the high-energy tail observed in $P(E)$ is an intrinsic feature of the present system.
This behavior is further supported by the temperature dependence of the transition 
rate  $T^E_{QT}$ that is analyzed in paper I.

\section{Kinetic energy distribution of adsorbate atoms: accommodation of normal and tangential energy components}\label{SecE}

In paper I~\cite{paper1} two types of states temporarily localized near the surface have been introduced. Typically, the quasi-trapped trajectories are characterized by a fast accommodation of the normal velocity component~\cite{smith1,smith2}. The equilibration times can be analyzed 
via the bounce number-dependent distribution functions for the parallel and normal kinetic energy components: 
$P(E^{\perp},n_b)$ and $P(E^{\parallel},n_b)$. 
 They can be directly related to the time-dependent distributions using the scaling factor $c$ from the time 
 dependence of the average bounce number, $\langle n_b \rangle(t) \approx c \cdot t$ (see paper I~\cite{paper1}). 
In the following, 
we use the value $t_0=10^3 \cdot a_0 \sqrt{m_{\mathrm{Ar}}/{E_\mathrm{h}}}=\unit[6.53]{ps}$ as a time unit, 
where $a_0$ is the Bohr radius, $m_{\mathrm{Ar}}$ denotes the atomic mass of the argon atoms, 
and $E_\mathrm{h} =\unit[27.211]{eV}$ is the Hartree energy. 

\subsection{Distribution function $P(E_k^{\perp},n_b)$ of the normal energy  component}
\begin{figure}
  \begin{center} 
  \hspace{-0.0cm}\includegraphics[width=0.49\textwidth]{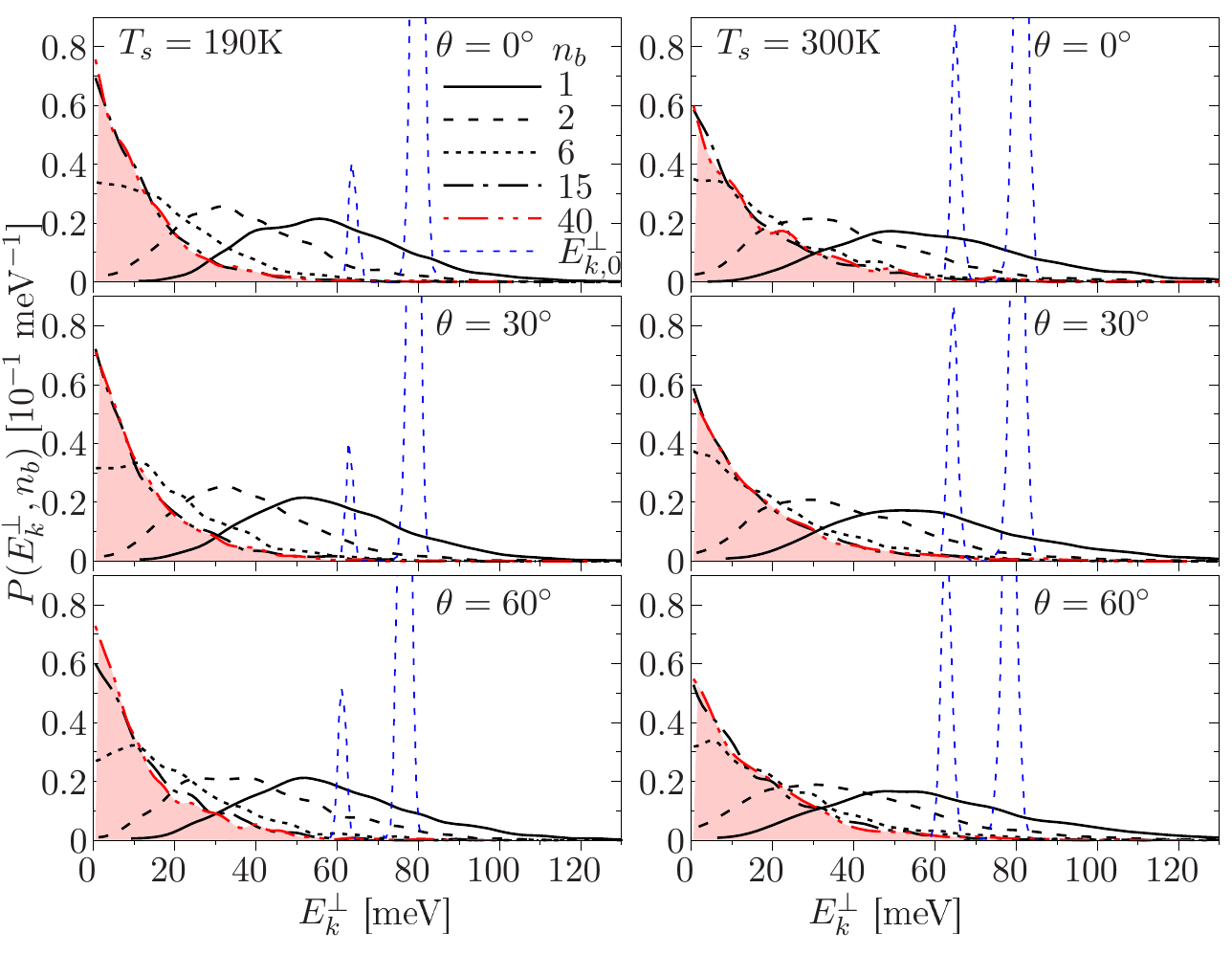} 
  \end{center}
  \vspace{-0.8cm}
  \caption{Distribution of the normal kinetic energy component $P(E_k^{\perp}, n_b)$ for different bounce numbers $n_b$ with $1\leq n_b \leq 40$. Two lattice temperatures ($T_s=190$ and $\unit[300]{K}$) and three incident angles ($\theta=0^\circ,30^\circ,60^\circ$) are compared. The incident kinetic energy for each $\theta$ is specified in figure~\ref{fig:dEfiT06n1}. The converged distribution ($n_b=40$) is shown by the filled pattern, and the dashed blue lines with the two sharp peaks represent the initial distribution just before the first bounce.}
  \label{fig:Ezdiffn}
 \end{figure}

First, we demonstrate the convergence of the normal component $E_k^{\perp}$ of the kinetic energy 
with increasing number of bounces $n_b$ 
in figure~\ref{fig:Ezdiffn}. Three incident angles $\theta$ and two lattice temperatures are compared. 
The incident kinetic energies are the same as in figure~\ref{fig:dEfiT06n1}.   
They are visible in the initial distribution $P(E_k^{\perp}, 0.4 t_0)$, 
represented by 
the dashed blue lines with two peaks at $E_k^{\perp}= 60$ and $\unit[80]{meV}$, 
which are computed 
at the time $t\sim 0.4 t_0 = \unit[2.6]{ps}$, just before the full energy exhibits the first jump due to the inelastic scattering. Note that the value $E_k^{\perp}$ is shifted from the incident kinetic energy $E_i^{\perp}$ by the depth of the physisorption well.
The particles reside at a distance of about $\unit[3-4]{\angstrom}$ from the surface and, hence, 
are strongly accelerated in the physisorption potential well. 
The initial distribution is very similar for the three angles. 
The two peaks correspond to the energy in the physisorption well at the two turning points, which corresponds to the atop and hollow site, respectively. 
This fact explains the separation of about $12$\,meV between the peaks. 
We conclude that the initial scattering conditions are very similar for all three cases. 
As a result, the subsequent evolution of the distribution function  $P(E_k^{\perp},n_b)$ with 
increasing 
$n_b$ follows the same trend and rapidly converges for $n_b> 6$. 
In particular, the distribution at $n_b=15$ practically coincides with the converged distribution (cf.\ the filled pattern, which refers to $n_b=40$ bounces). 

The left and right panels demonstrate the effect of thermal broadening specific to the given lattice temperature. For $T_s=300$K there is a reduction of the height of the distribution function at the origin, $P(E_k^{\perp}=0,n_b)|_{n_b\geq 15}$, an increased half-width and a broader high-energy tail. 
In conclusion, the MD simulations prove indeed a relatively fast convergence of the normal kinetic energy distribution function.

\subsection{Distribution function $P(E_k^{\parallel},n_b)$ of the in-plane energy component}
\begin{figure}
  \begin{center} 
  \hspace{-0.0cm}\includegraphics[width=0.49\textwidth]{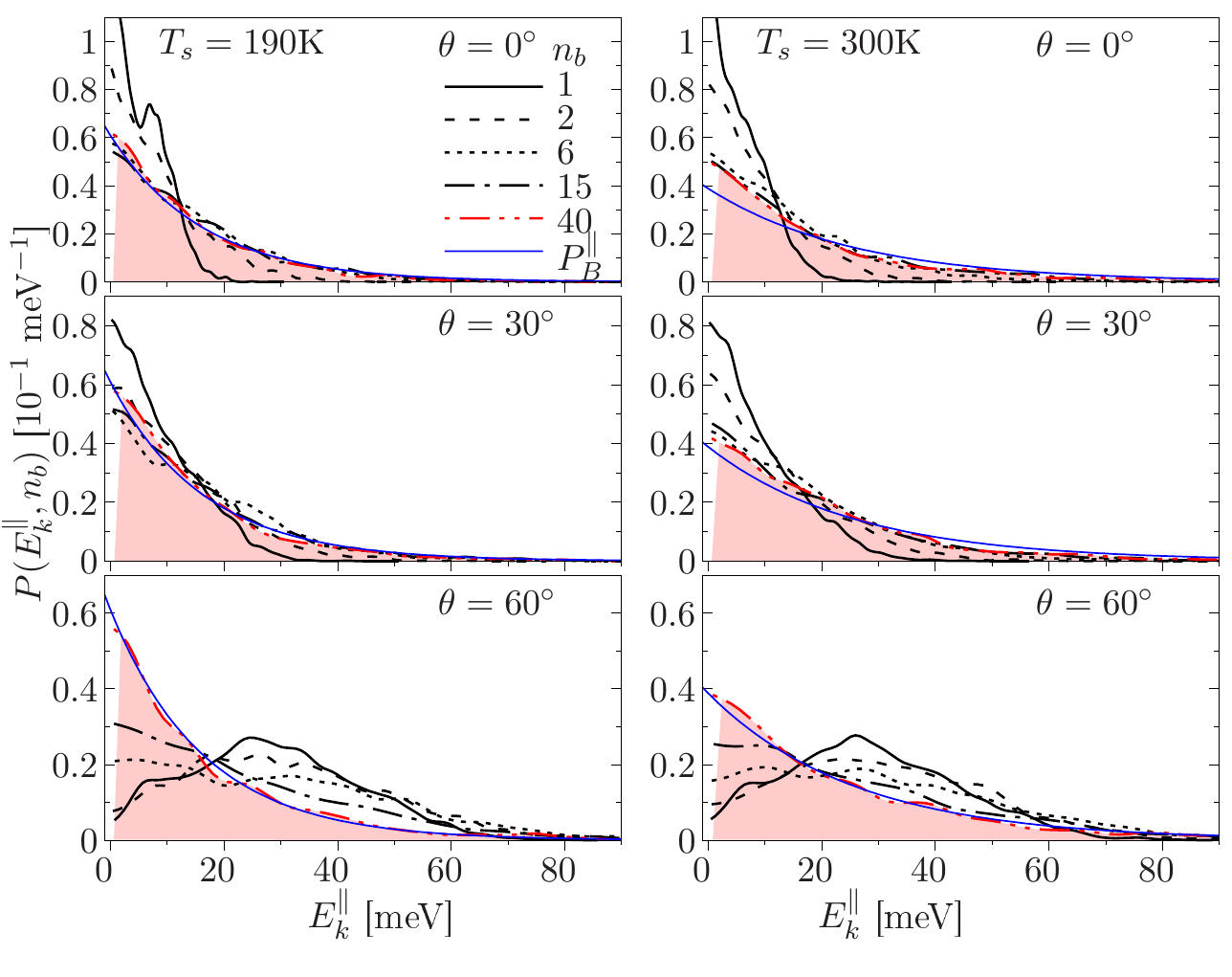} 
  \end{center}
  \vspace{-0.8cm}
  \caption{Distribution of the parallel kinetic energy component, $P(E_k^{\parallel}, n_b)$ for different bounce numbers $n_b$ with $1\leq n_b \leq 40$. 
  Two lattice temperatures ($T_s=190$ and $\unit[300]{K}$) and three incident angles ($\theta=0^\circ,30^\circ,60^\circ$) are compared. 
  The converged distribution ($n_b=40$) is shown by the filled pattern. The blue solid line is the prediction by the Boltzmann distribution, $P_B(E_k^{\parallel},T_s)$.}
  \label{fig:Exydiffn}
 \end{figure}

A similar analysis is performed for the parallel component 
$E_k^{\parallel}$ of the kinetic energy. The corresponding result is displayed in 
figure~\ref{fig:Exydiffn}. 
It is immediately evident that the initial distribution function $P(E_k^{\parallel},t)$ taken 
just before the first bounce 
at $t\sim 0.4 t_0$  significantly differs from the normal energy distribution shown in figure~\ref{fig:Ezdiffn}. 
It is peaked at the parallel kinetic energy in the incident beam, $E_{i}^{\parallel}=E_i \sin^2 \theta$. 
Up to some separation distance, this value is not influenced by the gas-surface interaction and, hence, the acceleration in the physisorption potential is irrelevant. As a result, a strong correlation with the initial energy is preserved for several first bounce events.

For the three angles  presented in figure~\ref{fig:Exydiffn}, 
the parallel components have been chosen accordingly as 
$E^{\parallel}_i=\unit[0]{meV}$ ($\theta=0^\circ$), $E^{\parallel}_i=\unit[3.98]{meV}$ ($\theta=30^\circ$), 
and $E^{\parallel}_i=\unit[27.26]{meV}$ ($\theta=60^\circ$). 
The requirement of a similar initial sticking probability for the  cases 
compared in section~\ref{stick} results in a relatively large parallel component 
for $\theta=60^\circ$,  whereas the normal components for the three cases stay comparable: 
$E^{\perp}_i=\unit[12.85]{meV}$ ($\theta=0^\circ$), $E^{\perp}_i=\unit[11.95]{meV}$ ($\theta=30^\circ$), and 
$E^{\perp}_i=\unit[9.07]{meV}$ ($\theta=60^\circ$). 

This choice of initial parameters explains to a large extent the slope of the distribution function after the first bounce. 
Due to a broad energy perturbation, $\Delta E\sim \unit[15]{meV}$, in a single inelastic scattering event, specified e.g.\ 
by the half-width of the distribution for $\theta=0^\circ$ at $n_b=1$, no qualitative difference is observed for the incident angles $0^\circ$ and $30^\circ$ when $E^{\parallel}_i \ll \Delta E$. For $\theta=60^\circ$ we observe a broad distribution of similar half-width $\Delta E$, centered around the initial energy $E^{\parallel}_i\sim 27$~meV. 
It gradually converges to the filled pattern ($n_b=40$) 
with the increase of the bounce number, 
but the convergence is much slower than for the normal 
component $E^{\perp}$ 
(cf.\ figure~\ref{fig:Ezdiffn}). 
Even for $n_b=15$ we still observe 
from figure~\ref{fig:Exydiffn}
that the population at low energies is reduced at the expense of the population at high energies. The distribution above $E^{\parallel}_i$ starts to decay significantly only for $6\leq n_b \leq 15$ or within $\unit[10]{ps}\leq t\leq \unit[20]{ps}$. 
On the same time scale, the fraction of the quasi-trapped states $N_Q(t)$ is reduced from $20\%$ to $8\%$. Hence, there is a direct relation between the $N_Q$ 
fraction and the population of high-energy states in  $P(E_k^{\parallel},n_b)$. 
The faster 
the depletion of $N_Q(t)$ due to the decay channels Q$\rightarrow$T and Q$\rightarrow$C 
takes place, the faster 
the convergence of the in-plane EDF to the quasi-equilibrium form is. 
The value of the transition 
rates  $T_{CQ(TQ)}$ (see paper I~\cite{paper1}), in its turn, 
crucially depends on the surface kinetics and the smoothness of the potential energy surface (PES) reconstructed by the Ar-Pt(111) interaction potential. 
For a flat PES the parallel kinetic energy component 
is 
less perturbed during a bounce event, and the 
conversion 
of the EDF to its stationary form takes longer. 

As an example, we consider the small incident angles $\theta=0^\circ$ and 
$30^\circ$. 
Here, the $N_Q$ fraction is small already after the first bounce (see figure~5 of paper I). 
As a result, at $t=\unit[20]{ps}$ (or $\langle n_b \rangle\sim 15$) its relative contribution to $P(E_k^{\parallel},n_b)$ 
is below $3\%$ in figure~\ref{fig:Exydiffn}, 
and the trapped states dominate. 
We can observe that the 
distribution function (DF) 
is close to its stationary form already after few bounces ($n_b \sim 6$) in this case.

Now the question arises why the distribution 
$P(E_k^{\perp},n_b)$ 
converges much faster than $P(E_k^{\parallel},n_b)$. 
The comparison of figures~\ref{fig:Ezdiffn} and~\ref{fig:Exydiffn} makes clear that 
$P(E_k^{\perp},n_b)$
contains the same amount of the quasi-trapped states as 
$P(E_k^{\parallel},n_b)$
in figure~\ref{fig:Exydiffn}.  
However, its 
transformation 
to the stationary form is 
reached already after few  bounces ($n_b\sim 10$).
  Moreover, the initial 
  DF   
  is peaked at a much higher energy ($60-80$~meV). 
  One main reason is the trapping condition $E_k^{\perp}+E_p<0$ 
  with the potential energy $E_p$, 
  which specifies an upper bound for the possible values of $E_k^{\perp}$. 
  Hence, all trajectories with the high-energy normal component vanish to the scattered fraction very fast.
\subsection{Relaxation of the in-plane 
EDF 
at low temperature}

\begin{figure}
  \begin{center} 
  \hspace{-0.0cm}\includegraphics[width=0.5\textwidth]{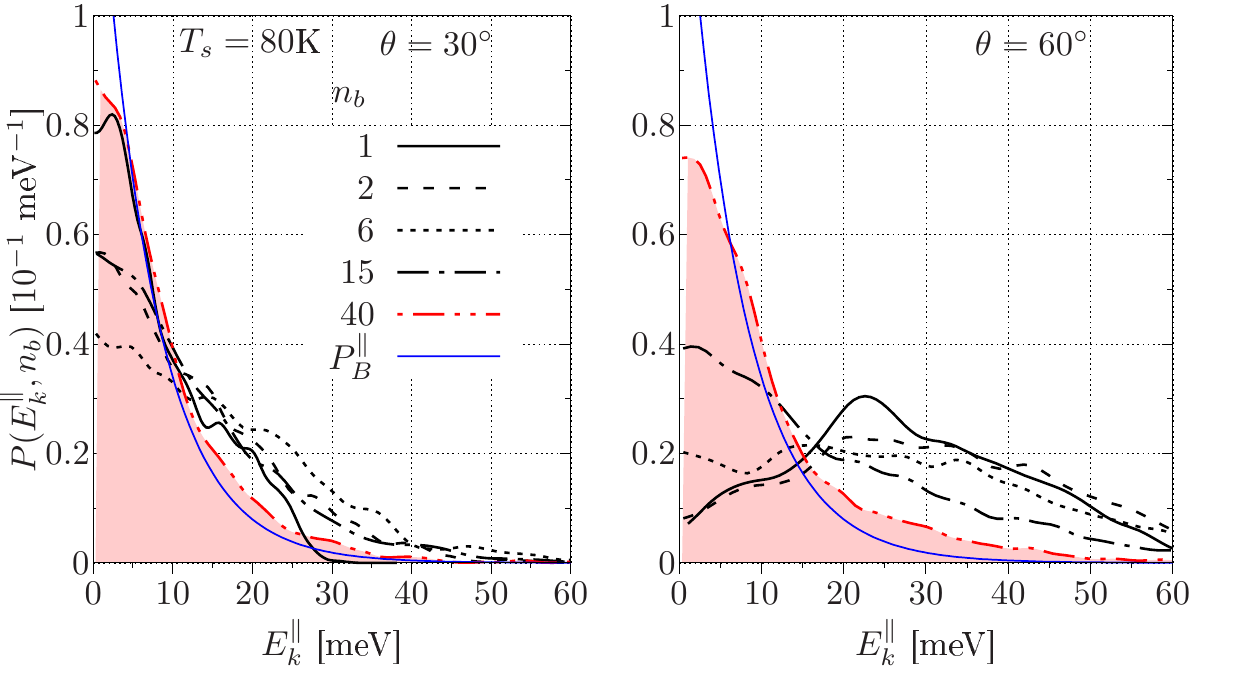} 
  \end{center}
  \vspace{-0.8cm}
  \caption{Distribution of the parallel kinetic energy component, $P(E_k^{\parallel}; n_b)$, at low lattice temperature, $T_s=\unit[80]{K}$, 
  at different bounce numbers 
  $1\leq n_b \leq 40$. Two incident angles ($\theta=30^\circ,60^\circ$) are compared. The converged distribution ($n_b=40$) is shown by the filled pattern. The blue solid line is the prediction by the Boltzmann distribution, $P_B(E_k^{\parallel},T_s)$.}
  \label{fig:ExydiffnT26}
 \end{figure}

In figure~\ref{fig:ExydiffnT26} we analyze the convergence 
behavior of the distribution of the parallel kinetic energy component
at low lattice temperature ($T_s=\unit[80]{K}$) and two incident angles ($30^\circ$, $60^\circ$). 
In contrast to the higher temperatures (cf.\ figure~\ref{fig:Exydiffn}), 
the convergence is not achieved at $n_b=15$. There are significant deviations from the DF at $n_b=40$. They can be better followed if the DF is compared with the Boltzmann prediction
for the lattice temperature of \unit[80]{K}, see Eq.~(\ref{modelB}) below.  Indeed, both distributions seem to converge to the thermodynamic prediction, when the adsorbate fully equilibrates with the thermal bath presented by the lattice atoms. 
There are still some discrepancies present at low and high energies. 
We can conclude that at such low temperatures the bounce number $n_b=40$ is not sufficient to reach a quasi-stationary distribution close to the Boltzmann prediction, even though the trapped states dominate the DF for both angles already at $t > 4.5 t_0$ ($\unit[30]{ps}$) 
according to the analysis presented in paper~I~\cite{paper1}. 

When comparing the results shown in figure~\ref{fig:ExydiffnT26}
with the  cases at higher temperatures displayed in  figure~\ref{fig:Exydiffn}, we 
actually observe that the latter DFs  have converged to the Boltzmann curve specified by some effective temperature $T^{\star}$ for all three angles and $n_b=40$. 
In the case of a lattice temperature  $T_s=\unit[190]{K}$ (left panel of figure~\ref{fig:Exydiffn}), 
we obtain the value $T^{\star}\sim \unit[150]{K}$. At $T_s=\unit[300]{K}$ the effect of the subthermal distribution 
becomes 
larger, 
and we find $T^{\star}\sim \unit[200]{K}$. 
Compared to the equilibrium Boltzmann prediction of $T^{\star}=T_s$, the measured DF shows an increased population of low-energy states. All this indicates that while the full convergence to an equilibrium function has been reached, the effective adsorbate temperature remains different from the lattice temperature with $T^{\star} < T_s$.
We note that a similar ``lag'' between $T^{\star}$ and $T_s$ has been measured for Ar on 2H-W(100) in~\cite{Rettner}
and for Ar on Pt(111) in~\cite{Hurst1985}. 
Our fitted values of $T^{\star}$ are found to be in very good agreement with these experimental data.

 \begin{figure}
   \begin{center} 
   \hspace{-0.0cm}\includegraphics[width=0.5\textwidth]{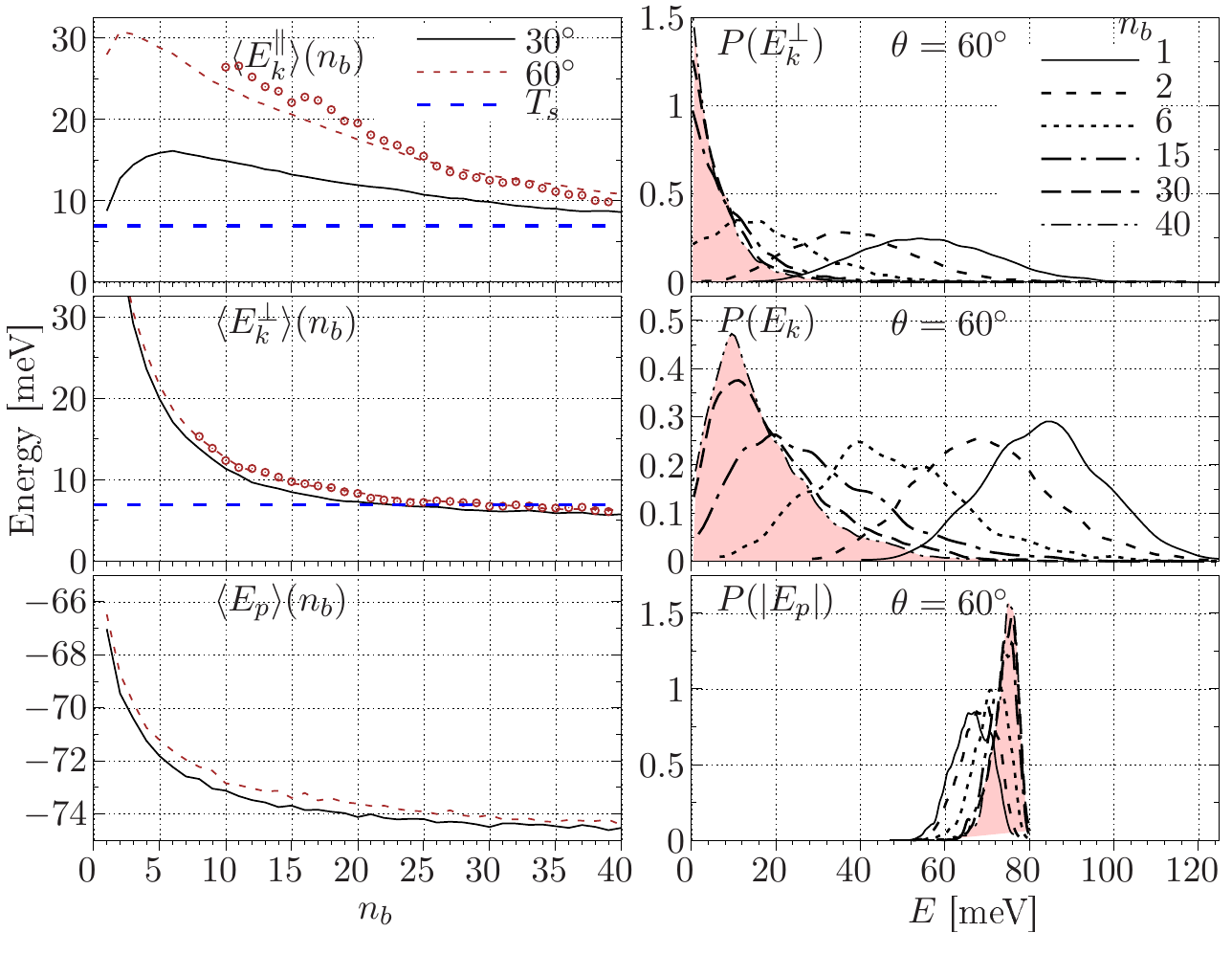} 
   \end{center}
   \vspace{-0.8cm}
   \caption{\textbf{Left:} Convergence of the parallel and normal components of the kinetic energy ($E_k^{\parallel}$, $E_k^{\perp}$)  and  potential energy ($E_p$) versus $n_b$, for the incident angles $\theta=30^{\circ},60^{\circ}$ and lattice temperature $T_s=80$\,K. \textbf{Right:} Associated evolution of the distribution functions of $E_k^{\perp}$, full kinetic energy ($E_k$) and $\abs{E_p}$.}
   \label{fig:expfitT26ag3060}
  \end{figure}   
  \begin{figure}
   \begin{center} 
   \hspace{-0.0cm}\includegraphics[width=0.5\textwidth]{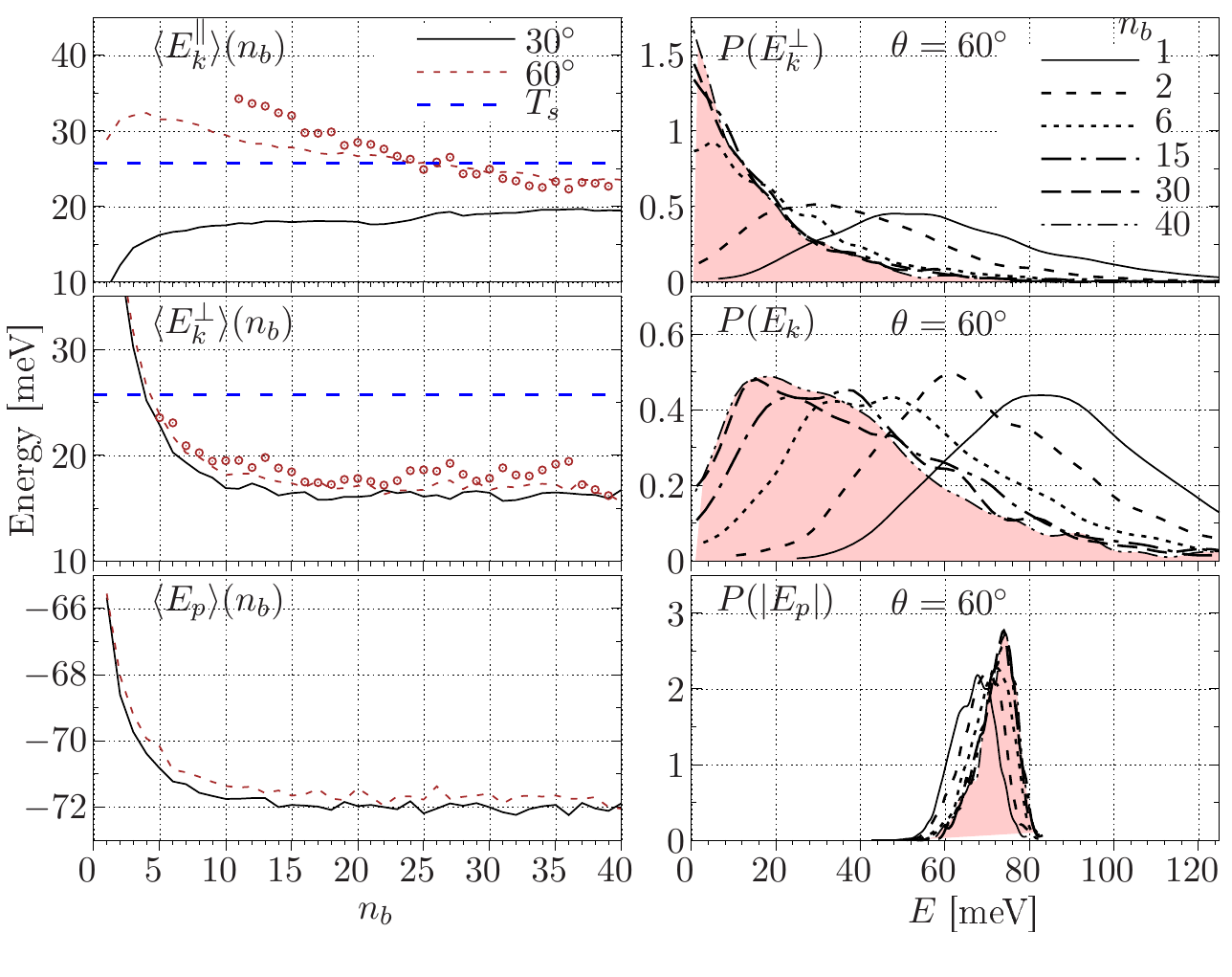} 
   \end{center}
   \vspace{-0.8cm}
   \caption{Same as in figure 9, but for the lattice temperature $T_s = 300$~K.}
   \label{fig:expfitT10ag3060}
  \end{figure}   
  
On the quantitative level, 
the convergence to equilibrium can be defined in terms of the effective temperature of the adsorbate. 
By analyzing the normal and tangential components (figures~\ref{fig:Ezdiffn},~\ref{fig:Exydiffn}, and~\ref{fig:ExydiffnT26}), 
we found that in most cases the DFs for $n_b\geq 9$ can be well 
fitted by  
the Boltzmann distribution 
\begin{equation}
  P(E^{(n)},n_b)  \approx  \frac{\pi}{T} \exp{\left[-E^{(n)}_{k}/T^{(n)}(n_b)\right]} 
  \label{modelB}
\end{equation}
with the temperature $T$ used as a time-dependent fit parameter and 
$n=(\perp,\parallel)$.
%
 Using this relation, 
 the average kinetic energy $\langle E_k^{(n)}\rangle(n_b)$, 
 given by 
  \begin{equation}
  \langle E_k^{(n)}\rangle(n_b) = \int \db E_{k}^{(n)} E_{k}^{(n)}\, P(E_{k}^{(n)},n_b) =T^{(n)}(n_b) \, ,
  \label{modelB1}
 \end{equation}
 should coincide with a true kinetic energy component 
 as soon as the model distribution~(\ref{modelB}) is close to the energy distribution determined by the MD simulations. 
 Such an analysis is presented 
 for the lattice temperatures $T_s=80$\,K and  $300$\,K in figures~\ref{fig:expfitT26ag3060} and \ref{fig:expfitT10ag3060}, respectively. 
 The two dashed lines in the two upper figures in the left panels show the decay of the MD kinetic energy versus the number of bounces. The open dots are the estimates based on Eq.~(\ref{modelB1}), where the effective temperature $T^{(n)}$ has been introduced as a fit parameter [shown are the results for $\theta=60^\circ$]. 

Note that in the Boltzmann case the average kinetic energy is a measure of the temperature: $\langle E_k^{(n)}\rangle(n_b)=T^{(n)}(n_b)$. Hence, both dependences demonstrate how fast the adsorbate effective temperature converges. As shown in figure~\ref{fig:expfitT26ag3060} for $T_s=80$\,K, the temperatures $T^{\parallel}$ and $T^{\perp}$ (and the kinetic energies $\langle E_k^{\parallel}\rangle$ and  $\langle E_k^{\perp}\rangle$ as well) 
approach  some effective temperature $T^{\star}$. It is comparable with the lattice temperature $T_s$, which is 
shown by the horizontal dashed line. In agreement with our previous discussion, we observe a much faster relaxation 
of the normal component. For comparison,  the convergence of the normal energy distribution, $P(E_k^{\perp})$, and of the total kinetic energy distribution, $P(E_k)$, versus $n_b$ can be followed in the right panels. The two lower figures illustrate 
the behavior of the average potential energy $\langle E_p \rangle(n_b)$ as a function of bounce number  
and of the corresponding distribution function $P(\abs{E_p})$ for selected $1 \le n_b \le 40$. 

The corresponding 
analysis for the higher lattice temperature, $T_s=300$\,K, 
is presented in figure~\ref{fig:expfitT10ag3060}. 
The main difference 
to the case with $T_s=80$\,K 
is that the converged effective temperature is $T^{\star} = \unit[200]{K}$ and  is well below $T_s$. 

The presented results bring us to the following conclusions: 
\begin{description}
  \item[i.] The Boltzmann distribution can approximately fit the simulation data for the distribution functions. 
  \item[ii.] In this case, the average kinetic energy is a measure of the effective adsorbate temperature $T^{(n)}$. 
  \item[iii.] In the long term, the value $T^{(n)}$ approaches some stationary value, typically, below the lattice temperature $T_s$.  
  \item[iv.] The MD simulations allow us to identify the equilibration times for different incidence conditions $\{E_i,\theta,T_s\}$.
\end{description}

\section{Sticking coefficient}\label{stick}
Now we focus on evaluation of the sticking coefficient and its dependence on the lattice temperature and the incidence conditions. Using our subdivision of particle trajectories, it is clear that both trapped and quasi-trapped particles contribute to this macroscopic observable defined as 
\begin{equation}
R_{st}(E_i,\theta,t) = N_T(E_i,\theta,t) + N_Q(E_i,\theta,t) \, .
\label{eq:sticking}
\end{equation}

The  contributions of the trapped and quasi-trapped states to the initial sticking coefficient after the first bounce, $n_b=1$, corresponding to  $t\lesssim 0.4 t_0$ $\approx 2.6$\,ps
are given in Table~\ref{tabs}. 
In particular, it is found that the contribution of trapped particles to $R_{st}$ 
decreases with increasing incident angle (and energy), while that of quasi-trapped particles increases at the same time.

\begin{table}[h]
\caption{Initial sticking probability and contributions of  trapped ($N_T$) and quasi-trapped ($N_Q$) particles, cf. Eq.~(\ref{eq:sticking}). Four incident angles and three lattice temperatures, $T_s$, are compared. The incident energies (in units of room temperature $T_r$) are $E_i/(k_BT_r/e_0)=0.5(0^\circ); 0.62(30^\circ); 0.84(45^\circ); 1.41(60^\circ)$.}
  \label{tabs}
 \begin{tabular}{c c c c c c}
 \hline
 \hline
$T_s$ & $\theta$ & $0^\circ$ & $30^\circ$ & $45^\circ$ & $60^\circ$ \\
 \hline
&$N_T$& 0.694   &  0.594 &0.398 &0.121\\
80\,K &$N_Q$& 0.095& 0.173 &0.317 &0.594\\
&$R_{st}$ & 0.789 & 0.767 & 0.715 & 0.715 \\
\hline
 &$N_T$& 0.570 &  0.522& 0.405 &0.187\\
190\,K &$N_Q$& 0.094& 0.154& 0.231 &0.466\\
&$R_{st}$ & 0.664 & 0.676 & 0.636 & 0.653 \\
\hline
&$N_T$& 0.531 & 0.493 & 0.417& 0.218\\
300\,K &$N_Q$&  0.074& 0.124&  0.194& 0.402\\
&$R_{st}$ & 0.605 & 0.617 & 0.611 & 0.620 \\
\hline
\hline
 \end{tabular}
 \end{table}

The temperature dependence of the initial sticking coefficient for normal incidence ($\theta=0^\circ$) 
can be compared with MD simulation results of Head-Gordon \etal~\cite{gordon}. 
Using  the condition $E_k^{\perp}+E_p<0$ 
as the trapping criterion, 
they report $R_{st}(80\,\mathrm{K})=0.72$, $R_{st}(180\,\mathrm{K})=0.62$ and $R_{st}(300\,\mathrm{K})=0.58$. 
Our values are systematically larger by $3\dots 10\%$. This discrepancy 
most probably originates from the different model 
used for the Pt-Pt and Ar-Pt interaction potentials.  
In Ref.~\cite{gordon} the platinum surface was  described by a harmonic nearest neighbor 
and next-nearest-neighbor interaction, and the Ar-Pt interaction by a 6-12 Lennard-Jones potential. 
Furthermore, the normal incident energy in Ref.~\cite{gordon} was $E_i=20.72$\,meV compared to $E_i=12.85$\,meV used in table~\ref{tabs}.

\subsection{Energy-resolved initial sticking coefficient ($n_b=1$). Comparison with experimental data}

During a collision event, the parallel momentum for a flat surface is conserved and does not play an 
important role in the trapping. Therefore, it can be assumed that 
the sticking probability depends solely on the energy exchange of the normal component 
in this ideal case,  and that it scales as $R_{st}(E_i,\theta)\sim R_{st}(E_k^{\perp})$ with $E_k^{\perp}=E_i \cos^2\theta$. 
If this assumption is valid, the measured sticking probabilities, 
obtained for a series of incident particle trajectories with the varying grazing angle $\theta$ and energy $E_i$, 
fit well with a common curve, when plotted as a function of $R_{st}(E_k^{\perp})$. 

Possible deviations from this ``normal scaling'' ($\cos^2\theta$) 
imply that the scattering surface is corrugated, e.g.\ due to the binary gas-surface atom interaction. 
The scaling $R_{st}(E_i\cos^n\theta)$ with $n <2$ indicates a dependence on the parallel momentum as well, 
whereas the limit $n=0$ 
refers to the dependence 
of $R_{st}$
on the total energy exchange between an atom and the surface.

   \begin{figure}
  \begin{center} 
  \hspace{-0.5cm}\includegraphics[width=0.45\textwidth]{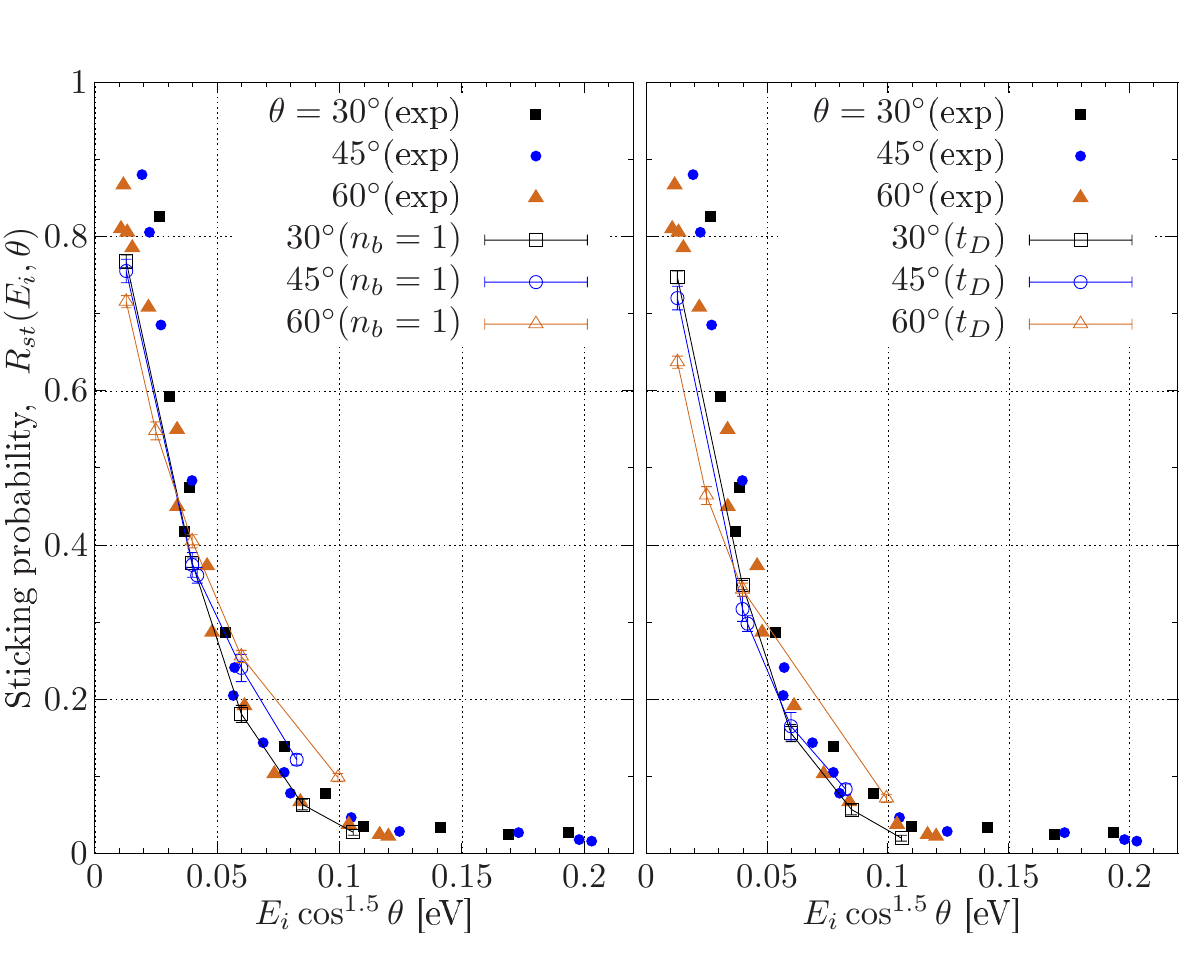} 
  \end{center}
  \vspace{-.70cm}
  \caption{Dependence of the sticking coefficient $R_{st}(E_i,\theta)$ for Ar on Pt(111) on the incident angle and energy for a lattice temperature $T_s=80$\,K. 
  $E_i$ is rescaled with $\cos^{1.5}\theta$. Full symbols: our simulations. Open symbols:  experimental data from Ref.~\cite{expPt}. \textbf{Left:} initial sticking coefficient after the first bounce ($n_b=1$). \textbf{Right:} sticking coefficient after a delay time $t_D=10$\,ps (see text).}
  \label{fig:Pt111-T80}
  \end{figure}

  \begin{figure}
  \begin{center} 
  \hspace{-0.5cm}\includegraphics[width=0.45\textwidth]{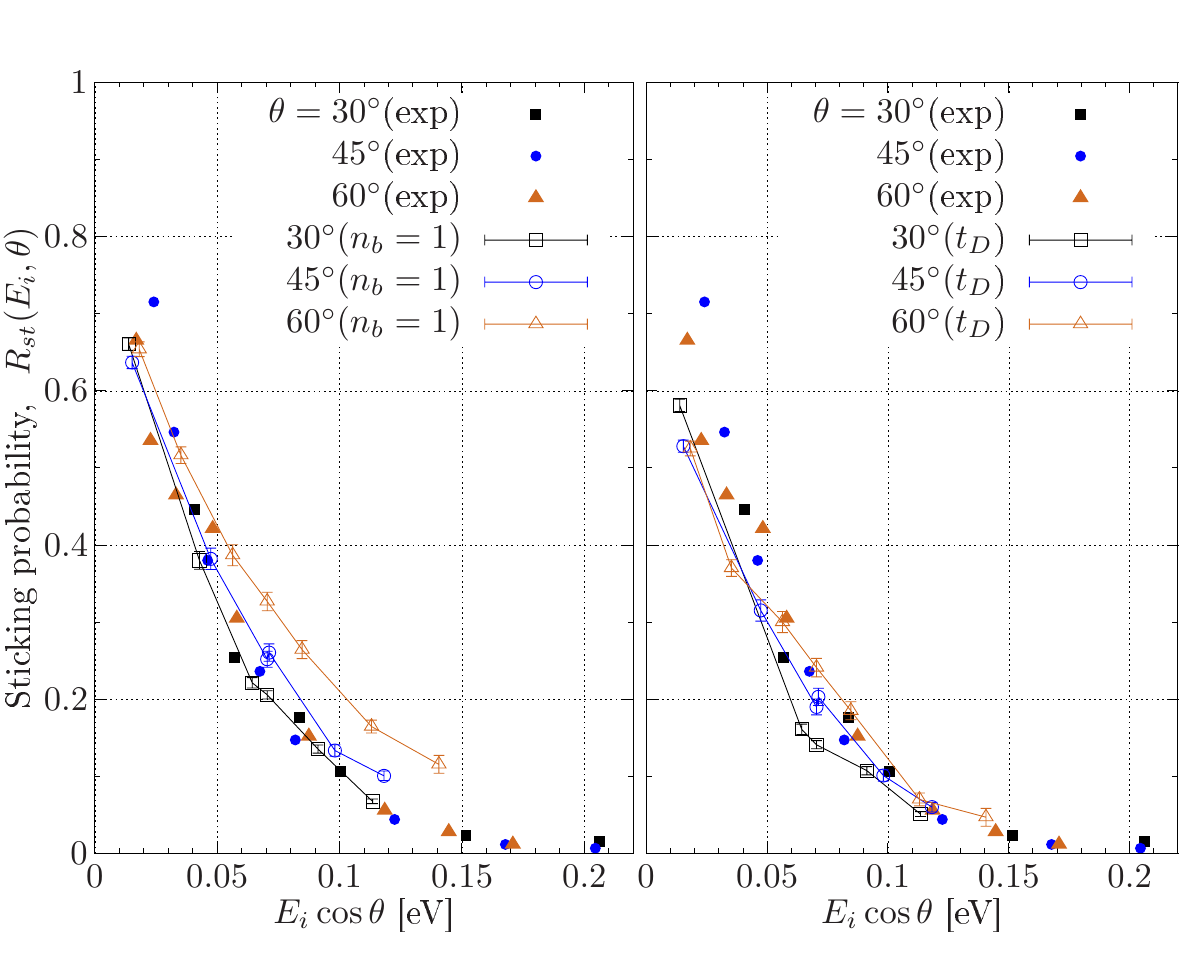} 
  \end{center}
  \vspace{-.70cm}
  \caption{Same as Fig.~\ref{fig:Pt111-T80}, but for $T_s=190$\,K and a different energy scaling according to $E_i\cos \theta$. The sticking probability measured with the delay time $t_D=10$\,ps (right panel) better agrees with the experiment~\cite{expPt}.}
  \label{fig:Pt111-T190}
  \end{figure}

Energy-resolved initial sticking coefficients 
obtained by our MD simulations 
at the lattice temperatures 
of  80, 190 and 300\,K are 
shown in figures~\ref{fig:Pt111-T80}, \ref{fig:Pt111-T190}, and \ref{fig:Pt111-T300}, respectively. 
In our simulations the 
sticking coefficients 
have been evaluated at three angles, $\theta=\{30^\circ,45^\circ,60^\circ\}$, as a function of the incident energy $E_i$. 
The incident atoms were initially randomly distributed within the area of the simulation cell to provide an average over the scattering events covering the whole area of the simulated sample.
The simulation results are compared with experimental data reported in~\cite{expPt}. 

The sticking probability decreases with the incident energy $E_i$ as well as with the rescaled energy $E_i\cos^n\theta$. 
Following the experimental data of Ref.~\cite{expPt}, we used the exponents $n=1.5$, $1.0$ and $0.5$ for the temperatures $T_s=80$\,K, $190$\,K, and $300$\,K, respectively. 
As a result of these scalings, most data points 
after the first bounce ($n_b=1$) 
shown on the left side of 
figures~\ref{fig:Pt111-T80}, \ref{fig:Pt111-T190}, and~\ref{fig:Pt111-T300} 
nearly fall on a single curve for different combinations of $E_i$ and $\theta$. 

Furthermore, 
much better agreement 
with the MD simulation results reported in~\cite{gordon} for $\theta=0^\circ$ 
can be resolved by the comparison with $R_{st}(E_i,\theta)$ reported in figures~\ref{fig:Pt111-T80}, \ref{fig:Pt111-T190}, and~\ref{fig:Pt111-T300} 
at the rescaled energy $E_i\cos^n\theta\sim 20$\,meV. 
Due to the expected universal dependence of $R_{st}$, the results reported for $\theta=30^\circ$ and $45^\circ$ 
should agree with the ones of Ref.~\cite{gordon} for $\theta=0^\circ$. 
This comparison shows good coincidence of the results and demonstrates that both models provide
a reasonable description of the gas-surface interaction energies and sticking in the thermal regime.

However, there still is a systematic discrepancy between our simulation results for the initial sticking coefficient 
after the first bounce and the experimental results, in particular, for the highest temperature (left side of figure~\ref{fig:Pt111-T300}). 
The origin of these deviations is analyzed in the following section, and a solution of this problem is presented. 

\subsection{Energy-resolved stationary sticking coefficient. Choice of the delay time $t_D$}
The best agreement with the experimental data and the assumed scaling is observed at $T_s=80$\,K (figure~\ref{fig:Pt111-T80}). 
As it is shown in figure~3 of paper I~\cite{paper1}, 
the scattered and the trapped states converge already after $t\sim 2 t_0 - 3 t_0$ corresponding to $13\dots 19$\,ps. 
The defined trapped fraction is preserved on a long time scale  
according to the estimated quasi-equilibrium desorption rates $T_{CT}$ and $T_{CQ}$ (see figure~9 in paper~I). 
Hence, there is less ambiguity in
the definition of the trapped states both experimentally and in the simulations.
\begin{figure}
  \begin{center} 
  \hspace{-0.5cm}\includegraphics[width=0.45\textwidth]{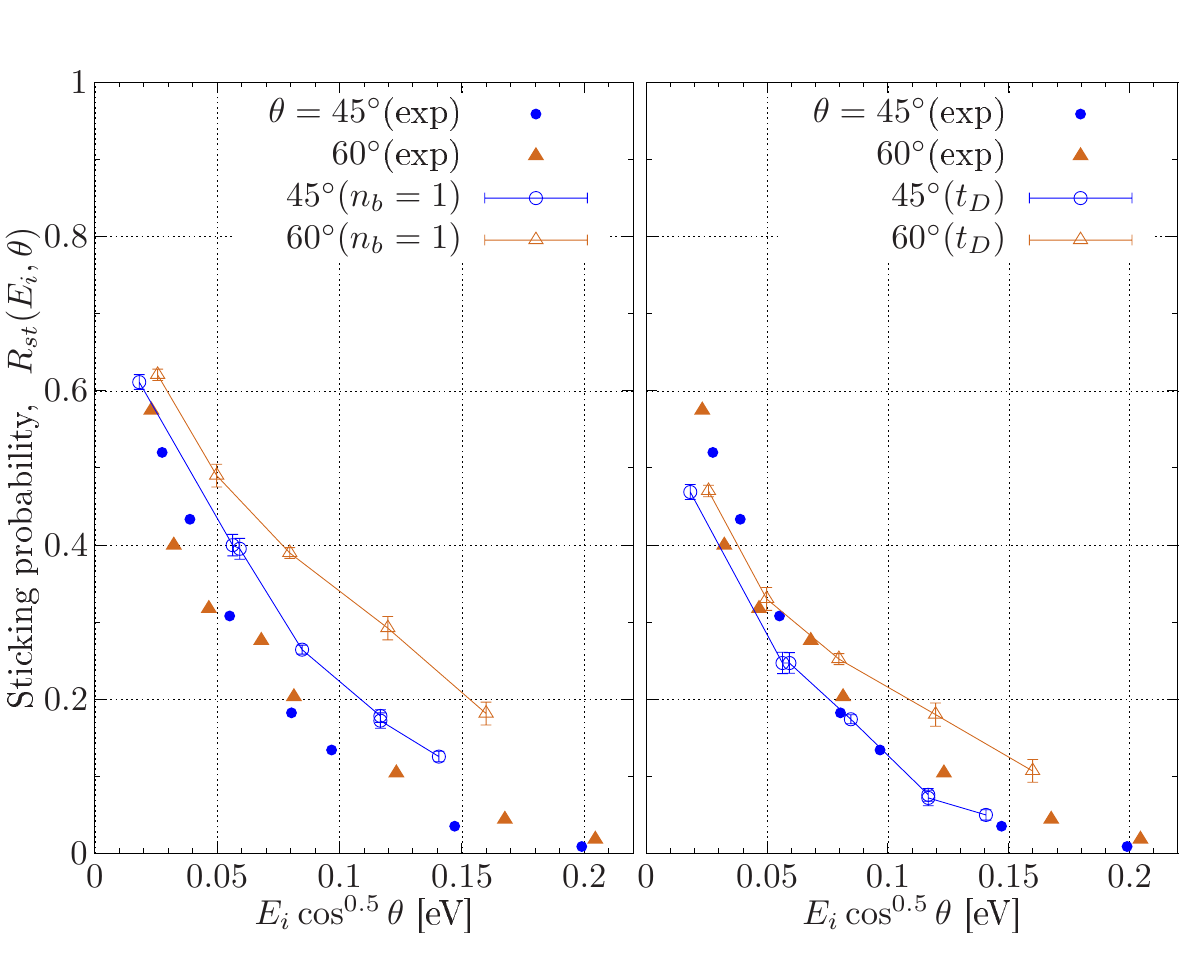} 
  \end{center}
  \vspace{-.70cm}
  \caption{Same as in Fig.~\ref{fig:Pt111-T190}, but for $T_s=300$\,K. The incident energy is rescaled as $E_i\cos^{0.5} \theta$.}
  \label{fig:Pt111-T300}
  \end{figure}

An uncertainty arises at the higher temperatures $T_s=\unit[190]{K}$ and $T_s=\unit[300]{K}$ 
(left sides of figures~\ref{fig:Pt111-T190} and~\ref{fig:Pt111-T300}). 
Once the desorption rates $T_{CT}$ and $T_{CQ}$ 
have finite values, 
the number of trapped states steadily decreases on a time scale of several picoseconds, $t\sim 15$\,ps 
(see figures~5 and~7 in paper I). 
Hence, the sticking fraction is dependent 
on a  delay time $t_D$ when the trapped particles can be desorbed at $t\leq t_D$ and counted as  scattered. 
It is difficult to specify this parameter in the experiment. 
Thus, the question arises how to compare experiment and theory (simulation), and for what specific value of $t_D$? 
In the following, we discuss this problem more in detail.

First, we note that  the assumed scaling $E_i(\theta)=E_i\cos^n\theta$ 
with the value $n$ used to fit the experimental data does not hold correctly for the MD data 
after one bounce 
for  higher temperatures 
at large energies and incident angles. Similar to the low temperature case ($T_s=80$\,K), 
we get  quantitative agreement with the experimental results for $\theta=30^\circ$. 
However, deviations appear systematically for $\theta=45^\circ$ and $60^\circ$. 
They increase at larger energies, i.e., for $E_i(\theta)> 0.05$\,eV in figure~\ref{fig:Pt111-T190} ($T_s=190$\,K) 
and for $E_i(\theta)> 0.025$\,eV in figure~\ref{fig:Pt111-T300} ($T_s=300$\,K). 
The MD results (left panel) correspond to the initial trapped fraction and, hence, 
they should provide an upper bound for the trapped fraction measured with some time delay (experimental data). 
This is indeed the case for both the temperatures, and the difference between the predicted upper bound and the experimental data increases with $\theta$ and $E_i(\theta)$. 
In particular,  the deviations at a fixed value of $E_i(\theta)$ are always larger for $\theta=60^\circ$ than for $\theta=45^\circ$. 
Note that the absolute energy $E_i$ is larger for $\theta=60^\circ$ due to the scaling factor $\cos^n \theta$ ($n\geq 1$). 
If we now refer to the desorption rate $T_{QT}(T,E)$ (see figure~9 in paper~I), 
which corresponds to the first stage in the desorption process 
$\mathrm{T} \rightarrow \mathrm{Q} \rightarrow \mathrm{C}$, 
we find that its value increases with energy. Hence, the desorption fraction accumulated during a fixed time delay 
becomes larger for $\theta=60^\circ$ than for $\theta=45^\circ$. 
In our opinion,
this is the main reason for the increase of deviations at larger $\theta$ and $E_i(\theta)$.

In order to make a proper comparison, we propose to introduce a fit parameter $t_D$, i.e., a delay time 
specific to a given experiment and related to a setup used to identify the desorbed (or the trapped) fraction. 
The sticking probability defined in the experiment can be written as
\begin{eqnarray}
 R_{st}^{\text{exp}}(E_i,\theta) &\approx & R_{st}^{\text{MD}}(E_i,\theta,t_D)=1- N_C(t_D) \, , \label{rtmd} \\
 N_C(t_D) &=& N_C(t_1)+\db N_C(t_1,t_D) \, , \label{eq:NCtD} \\
 \db N_C(t_1,t_D) &=&  \int_{t_1}^{t_D} \db t\, \left[T_{CQ}(t)N_Q(t) \right. \nonumber \\
   &&  
\left. + T_{CT}(t)N_T(t)\right] \, ,
\label{eq:NCt0tD}
\end{eqnarray}
where $N_C(t_1)$ is the fraction of atoms which experience a direct inelastic scattering to continuum by the first reflection from the surface (at some average time $t_1$), 
$\db N_C(t_1,t_D)$ is the contribution from the atoms, which have been initially trapped but desorbed during the time interval $t_1<t\leq t_D$. 
Equation~(\ref{eq:NCt0tD}) specifies the relation to the 
transition rates from the quasi-trapped and trapped states to the continuum.

For the present analysis, we defined $t_D \sim 1.5 t_0 \sim 10$\,ps 
empirically from a fit of a single MD data point to a single experimental point, taken at the specific lattice temperature, $T_s=190$K, the incident angle, $\theta=45^\circ$, and the energy,  $E_i(\theta)\sim 0.12$~eV, i.e.,  
\begin{eqnarray}
R_{st}^{\text{exp}}(E_i,\theta,T_s)=R_{st}^{\text{MD}}(E_i,\theta,t_D,T_s).  
\end{eqnarray}

The sticking probability, defined via Eq.~(\ref{rtmd}) with the fixed delay time $t_D=1.5 t_0$, is now compared with the experimental data. 
This comparison is displayed in the right panels of figures~\ref{fig:Pt111-T80}, \ref{fig:Pt111-T190}, and~\ref{fig:Pt111-T300}. 
In general, we observe  much better agreement taking into account that the value $t_D$ was not adjusted separately in each case. The results for $T_s=80$\,K remain practically unchanged, as  there is no noticeable thermal desorption on the time scale $t\leq t_D$. 
Much better agreement is now observed for $\theta=45^\circ$ and $60^\circ$ at $T_s=190$\,K 
(figure~\ref{fig:Pt111-T190}) 
and for $\theta=45^\circ$ at $T_s=300$\,K (figure~\ref{fig:Pt111-T300}). 
The remaining discrepancy for $\theta=60^\circ$ and $T_s=300$\,K can be further reduced by choosing $t_D \sim (2 \dots 3) t_0$.

Now we argue that a proper choice of $t_D$ should be the quasi-equilibration time $t^E$. This is related with the procedure, how the trapped fraction is determined experimentally. Typically, the procedure is well defined at low lattice temperatures ($\sim 80$\,K). 
The detected kinetic energy distribution in the angle-resolved reflected flux shows a bimodal distribution with one sharp and a second broad maxima. The first 
maximum is related with the specular reflection (corresponding to one-two bounces with the surface), and its position shows a strong correlation with the incidence conditions. In contrast, the second maximum stays independent, and, therefore, is related with the trapping-desorption contribution of particles which are temporarily trapped near the surface and partially equilibrate. The relative contribution of both varies strongly with the detection angle~\cite{Ar,collins,Mans}. By choosing the detection angle far from the incident one (where the specular reflection dominates), the measured intensity is mainly 
due to the trapped and quasi-trapped states. By analyzing the temporal evolution of the energy spectra related with the trapping-desorption fraction, one can study its convergence to a quasi-stationary form, and extract the  quasi-equilibration time $t^E$. Then, the integrated intensity taken at $t\geq t^E(=t^D)$  can be used as a definition of the trapped fraction and the sticking coefficient. The resolved characteristic time $t^E$ can be compared versus the MD simulations. From the theoretical side, the value 
of $t^E$ can be  determined independently using several of the  criteria mentioned in the previous sections. This includes the convergence of the kinetic energy distributions, the convergence of the average kinetic energy, and the time-dependence of the transition rates, cf.\ paper~I~\cite{paper1}.

\section{Conclusion and outlook for plasma-surface interaction}\label{con}

One of the fundamental aspects of rare gas-metal surface scattering experimental studies is the phenomenological division~\cite{Ar,N2,Xe} of this process into two distinct channels: the direct inelastic and the trapping desorption one. At shorter times a beam energy-dependent 
peak is observed that is attributed to direct inelastic scattering.
On the other hand, at longer time scales, the energy (time-of-flight) spectra exhibit a second, beam energy-independent peak that is attributed to the trapping desorption scattering. It shows nearly Maxwellian behavior at the surface temperature $T_s$. These experimental findings get a firm conformation from the present MD simulations, where the adsorbate equilibration kinetics has been studied in detail. By performing relatively long simulations we are able to distinguish and quantify two characteristic regimes. 
 At shorter times, i.e., within the first $5-10$ bounces with the surface, corresponding to a few picoseconds, the energy distribution functions of the adsorbate atoms exhibit fast changes. Then, at later times they converge to a quasi-stationary form, which stays practically unchanged until the end of the simulations. 
    This fact confirms that the phenomenological assumption of the two types of scattering processes is indeed valid.
 However, we also found that the corresponding time scales, 
 for which these processes can be well separated, strongly depend 
 on the incidence conditions of the atoms and on the lattice temperature. It was one of the goals of the present study to make quantitative predictions on such relevant time scales. The results can be used for a better interpretation of  experimental energy-resolved (time-of-flight) spectra.

The combination of the present MD simulations, based on microscopic gas-surface and binary atom-atom interactions, with the rate equation model, described in more detail in paper I~\cite{paper1}, allows us to directly analyze the temporal evolution of the trapping-desorption fraction. This quantity is difficult to access experimentally. 
Therefore, our combined  model allows for valuable predictions of the time scales when the behavior of the trapping-desorption fraction changes from a non-stationary  one to a quasi-equilibrium state. This was demonstrated on the example of several macroscopic observables.

The temperature dependence of the 
ELDF has also been analyzed. The half-width of the ELDF exhibits a linear scaling with the lattice temperature $T_s$. In the quasi-equilibrium regime ($t> t^E$) the ELDF is symmetric with respect to zero energy. This justifies that  
the probabilities of excitation and de-excitation become equal due to the energy exchange with the lattice. 
In this case the adsorbate average kinetic energy fully accommodates to some effective temperature $T^{\star}$ which is comparable to $T_s$. 
In general, we found that the accommodation time increases with the population of the quasi-trapped states $N_Q$. 
The latter increases with the angle of incidence of the gas atoms.

The convergence of the parallel and normal components of the kinetic energy has been studied for several typical incident angles and $T_s$ values. 
In agreement with  previous studies~\cite{Auerbach,smith1,smith2}, the parallel component always shows a slower convergence. The time-resolved kinetic energy distribution functions can be fitted by a Maxwellian distribution at some effective temperature, which typically shows a (subthermal) ``lag'' with respect to the lattice temperature. 

Finally, we checked the quality of our model by comparison with experimental data of the sticking coefficient~\cite{expPt}. 
While the  experimental results for low lattice temperature ($T_s=80$\,K) have been accurately reproduced, 
systematic discrepancies are observed for higher temperatures ($T_s=190$\,K and $300$\,K). 
To get  better agreement with the experimental data, we proposed to introduce a delay time, $t_D$, which takes into account the fraction of atoms which have been initially trapped, but have desorbed at $t<t_D$ and counted in the experiment as scattered. The empirical choice of this fit parameter, as $t_D\sim t^E$ ($10-30$\,ps), i.e., of the order of the quasi-equilibration time, brings our MD data into much better agreement with the experiments. This fact most likely explains similar discrepancies that were observed in other numerical simulations as well, see Refs.~\cite{smith1,smith2,Kul}

In summary, we studied the trapping-desorption processes, being the dominant mechanism for the scattering of rare gas atoms from metal surfaces, 
for the case of thermal and subthermal impact energy.  The simulation results generally agree with the experimental data and provide estimates of the characteristic quasi-equilibration times, the energy loss and kinetic energy spectra. 
The effects of the lattice temperature and incidence conditions have been analyzed in detail. Our theory and simulations are expected to be a useful tool for the analysis of atomic and molecular scattering, when the trapping-desorption mechanism or the initial trapping fraction (relative to the direct-scattered fraction) dominate.

Our results are expected to be of particular interest for low-temperature  plasmas 
at low pressure 
for which the considered energy range is of relevance. Due to the low degree of ionization the behavior of neutral gas atoms in the vicinity of a solid surface is crucial both, for fundamental understanding and for many applications, for a recent overview see Ref.~\cite{plasma-road-map-17}. First of all, our results for the sticking probability can be used to compute improved sticking coefficients of rare gas atoms on a metal surface for typical 
low-temperature plasma conditions. To this end, the present results for monoenergetic atoms can be averaged with the proper energy and angle distribution of atoms. Moreover, the same procedure can be applied to nonequilibrium conditions in the plasma sheath which are resulting e.g. from charge-exchange collisions. 

Second, our results could serve as an input for particle based simulations, such as 
particle-in-cell, 
molecular dynamics or kinetic Monte Carlo simulations, see Ref.~\cite{bonitz_cpp_12} and references therein. 
The present energy and angle resolved sticking probabilities  should allow for much more accurate simulation of surface processes that have predictive capability.

Third, the present method can be easily extended to other material combinations, in particular for metal surfaces, provided suitable \textit{ab initio} pair potentials or force fields are available that can be used as input for MD simulations.

Fourth, the capability of our combined MD-rate equation approach to perform accurate long-time simulations can be a valuable starting point to study slow processes such as surface modification due to sputtering or film growth. For example, our approach is not limited to an ideal metal surface but could be equally applied to a realistic corrugated surface that includes steps or defects.

Finally, our approach should, in principle, also allow for an \textit{ab initio} study of adsorption processes. 
Here, the main difference compared to the present study 
is 
that sticking of atoms 
is not
independent of the state of the surface. 
Instead, it is possible to track the pre-existing adsorbed atoms and include their influence on the sticking of atoms that arrive at a later  point in time. 
The present MD simulations can be easily extended by including proper atom-atom pair potentials. At the same time, the present rate equation model~\cite{paper1} 
can be straightforwardly extended to inhomogeneous systems and to the computation of atomic pair distribution or triple correlation functions~\cite{thomsen_pre_15}. This provides a suitable route to the computation of structural properties of surfaces in low-temperature plasmas and to the comparison with experimental data. 

%


\end{document}